\documentclass[usenatbib]{mnras}

\usepackage[T1]{fontenc}
\usepackage{ae,aecompl}

\usepackage{graphicx}	% Including figure files
\usepackage{amsmath}	% Advanced maths commands
\usepackage{amssymb}	% Extra maths symbols
\newcommand{\Zabs}{$Z_{\textrm{abs}}$}
\newcommand{\Zem}{$Z_{\textrm{emiss}}$}
\newcommand{\Zmax}{$Z_{\textrm{max}}$}
\newcommand{\logZabs}{$\log(Z_{\textrm{abs}} / Z_\odot )$}
\newcommand{\logZem}{$\log(Z_{\textrm{emiss}} / Z_\odot )$}
\newcommand{\Hii}{H\textsc{ii} }

\title[The internal Z-distribution of GRB host galaxies]{The internal metallicity distributions of simulated galaxies from EAGLE, Illustris, and IllustrisTNG at $z=1.8-4$ as probed by Gamma Ray Burst hosts}

\author[Metha, B., Trenti. M.]{
Benjamin Metha$^{1,2}$\thanks{\hbox{methab@student.unimelb.edu.au}}, Michele Trenti$^{1,2}$
\\
$^1$School of Physics, The University of Melbourne, VIC 3010, Australia\\
$^2$Australian Research Council Centre of Excellence for All-Sky Astrophysics in 3-Dimensions, Australia
}

\date{Accepted XXX. Received YYY; in original form ZZZ}

\pubyear{2019}

\begin{document}
\label{firstpage}
\pagerange{\pageref{firstpage}--\pageref{lastpage}}
\maketitle

\begin{abstract}
Massive stars are thought to be progenitors of Long Gamma Ray Bursts (GRBs), most likely with a bias favouring low metallicity progenitors. Because galaxies do not have a constant metallicity throughout, the combination of line-of-sight absorption metallicity inferred from GRB afterglow spectroscopy and of host galaxy global metallicity derived from emission lines diagnostics represents a powerful way to probe both the bias function for GRB progenitors, and the chemical inhomogeneities across star forming regions. In this study, we predict the relationship between \Zabs\ and \Zem\ using three different hydrodynamical cosmological simulations: Illustris, EAGLE, and IllustrisTNG. We find that while the qualitative shape of the curve relating emission versus absorption metallicity remains the same, the predicted relationship between these two observables is significantly different between the simulations. Using data for the host galaxy of GRB121024A for which both \Zabs\ and \Zem\ have been measured, we find marginal support for the Illustris simulation as producing the most-realistic internal metallicity distributions within star-forming galaxies at cosmic noon. Overall, all simulations predict similar properties for the bulk of the GRB host galaxy population, but each has distinct features in the tail of the \Zabs-\Zem\ distribution that in principle allow to discriminate between models if a sufficiently large sample of observations are available (i.e. $N\gtrsim 11$ on average). Substantial progress is expected in the near future, with upcoming JWST/NIRspec observations of 10 GRB host galaxies for which absorption metallicity from the afterglow spectra exists.

\end{abstract}


%%%%%%%%%%%%%%%%%%%%%%%%%%%%%%%%%%%%%%%%%%%%%%%%%%%%
%%%%%%%%%%%%%%%%%%%%%%%%%%%%%%%%%%%%%%%%%%%%%%%%%%%%
%%%%%%%%%%%%%%%%% BODY OF PAPER %%%%%%%%%%%%%%%%%%%%
%%%%%%%%%%%%%%%%%%%%%%%%%%%%%%%%%%%%%%%%%%%%%%%%%%%%
%%%%%%%%%%%%%%%%%%%%%%%%%%%%%%%%%%%%%%%%%%%%%%%%%%%%

\section{Introduction} \label{sec:intro}

Over the past few decades, significant progress has been made in understanding how galaxies form and evolve over time. 
Thanks to recent advancements in numerical methods and an improved understanding of galaxy-scale physics, several different high-resolution cosmological hydrodynamical simulations of sufficiently large volumes of the Universe exist, all of which reproduce to good accuracy the observed properties of galaxy populations in the local universe (see e.g. \citealt{Vogelsberger+20} for a review). Interestingly, it is often the case that these simulations rely on very different subgrid physical prescriptions, yet agree in the predicted bulk properties of galaxy populations \citep{Naab+Ostriker17}. An opportunity to discriminate among different subgrid physical prescriptions that give rise to similar global properties of galaxies is thus to investigate predictions of the internal structures of galaxies.

One way in which this can be done is by comparing the internal gas-phase metallicity distributions of simulated galaxies to observations.\footnote{Hereafter, we will refer to gas-phase metallicity simply as ``metallicity".} Metallicity (denoted by the symbol $Z$) is a fundamental property of galaxies. Metals (elements heavier than Helium) are produced in stars \citep{Burbidge+57} and released into the interstellar media (ISM) of galaxies by supernovae. The total metallicity of a galaxy, therefore, is related to the total amount of stars that a galaxy has formed over its lifetime \citep[e.g.][]{Maiolino+Mannucci19}.

Observations have shown that galaxies do not have constant metallicities throughout. For star-forming galaxies in the local universe, the central regions of galaxies generally exhibit more metal enrichment than their outskirts \citep[e.g.][]{Searle71, Shaver+83, VilaCostas+92, Berg+13, Berg+20, Ho+15, Belfiore+2017, Sanchez20}.  The average rate at which metallicity changes with the distance from a galaxy's centre is described as a \textit{metallicity gradient}, with a positive gradient meaning lower central metallicity.  

Negative metallicity gradients are theoretically well understood. Analytical studies have been able to explain the presence of negative radial metallicity gradients in galaxy discs by accounting for the difference in time scales between gas flows and star formation \citep{Tinsley80}, assuming either an exponential star formation profile within galaxy discs \citep[e.g][]{Edmunds+Greenhow95, Mo+98}, or that galaxies grow over time due to the accretion of gas onto their outskirts, triggering star-formation at larger galactic radii as time goes on (the \textit{inside-out model of galaxy formation}, see e.g. \citealt{Boissier+Prantzos99, Pilkington+12}). Negative metallicity gradients have also been seen in numerical simulations of galaxy formation. As an early example, \citet{Churches+01} predicted that negative metallicity gradients establish quickly from the combination of an exponential gas profile \citep{Freeman70} with a Schmidt star formation law \citep{Schmidt59, Kennicutt89}. Modern studies have been conducted comparing the metallicity gradients predicted for galaxies in state-of-the-art simulations to observations, finding that these simulations produce fairly realistic gradients for local galaxies \citep{Hemler+21, Tissera+22}, and allowing connections to be drawn between metallicity gradients and cold gas accretion, stellar feedback, and galaxy mergers \citep[e.g.][]{Porter+22}. 

At higher redshifts ($z \sim 2-4$), the internal metallicity structure of star-forming galaxies becomes more complicated. Often the metallicity structure is asymmetric, clumpy, and irregular \citep[e.g.][]{Guo+12, Johnson+17, Wang+20, Wang+22}. When a linear metallicity profile is fit, these galaxies may show flattened gradients \citep[e.g.][]{Kewley+06, Rupke+10, Wuyts+16} or inverted positive gradients \citep[e.g.][]{Cresci+10, Queyrel+12, Wang+22}, which can be explained through either galaxy-galaxy interactions \citep[e.g.][]{Tissera+16}, stronger feedback in low-mass starbursting systems disrupting gradients \citep[e.g.][]{Ma+17}, or the funnelling of cold gas into the central regions of these galaxies \citep[e.g.][]{Dekel+09a, Dekel+09b, Stott+14, Ellison+18}. These inverted gradients have also been seen in cosmological simulations at high redshift \citep[e.g][]{Hemler+21, Tissera+22}. However, the two-dimensional metallicity profiles of these galaxies often show large azimuthal asymmetries, raising the question whether describing such galaxies with a simple metallicity gradient model is appropriate \citep{Curti+20, Florian+21}. Whether or not the detailed chemical structure of such galaxies can be predicted from any subgrid models of galaxy evolution is yet to be determined. Observations of the internal metallicity structures of these galaxies are challenging due to their large distances, small sizes, intrinsic faintness, and the presence of star-forming clumps with uncharacteristically low/high metallicities \citep{Yuan+13, Carton+17}. However, it is possible to indirectly compare the internal metallicity structures of real and simulated high-redshift galaxies by assessing the relationship between emission metallicities (\Zem) determined using well-calibrated diagnostics and emission-line ratios, and absorption metallicities (\Zabs) associated with long gamma-ray bursts (GRBs).

Strong emission-line diagnostics are widely used to determine the gas-phase metallicities of galaxies \citep{Maiolino+Mannucci19, Kewley+19}. These diagnostics allow ratios of bright emission lines within a galaxy to be converted into metallicity determinations, and may be calibrated either to theoretical models of H\textsc{ii} regions \citep[e.g.][]{Kewley+Dopita02, Dopita+16}, or empirical metallicity observations determined directly by measuring the temperatures of interstellar gas \citep[e.g.][]{Pilyugin+Thaun05, Pilyugin+Grebel16}. Regardless of the diagnostic used, such emission-line based metallicity measurements (hereafter \Zem) will be necessarily biased to the brightest regions of the galaxy, in which young stars have recently been formed.

An alternative way to determine the metallicity of a high-redshift galaxy is by using absorption-based spectroscopy (\Zabs\ hereafter). This method relies on measuring the absorption spectrum of a bright background source, such as a quasar or a GRB, as it passes through the galaxy. Unlike \Zem, \Zabs\ is not biased to regions of higher star formation, as the absorbing abilities of gas does not depend on its ionisation state.
However, these absorption methods only measure the metallicity along the line-of-sight between the observer and the source of the emission.

Differences have been found between \Zabs\ and \Zem\ in quasar-absorbing systems. Invariably, \Zem\ is greater than \Zabs\ \citep[e.g.][]{Chen+05, Peroux+14}, but this difference can be largely explained by the large distances between quasar lines-of-sight and the galaxy centre, together with a negative metallicity gradient model \citep[e.g.][]{Christensen+14, Rahmani+16, Krogager+20}. For the local dwarf galaxy SBS 1543+593, metallicities inferred from the absorption spectrum of a quasar that passes through the galactic disk are consistent with emission-line metallicities inferred from H\textsc{ii} region spectroscopy \citep{Schulte-Ladbeck+04, Schulte-Ladbeck+05}. All of this implies that for quasar absorbing systems, \Zem\ and \Zabs\ are consistent with each other, implying that there is no large intrinsic bias between \Zem\ and \Zabs, and any observed difference should come from physical properties of the galaxies observed.

For GRB host systems, we do not expect \Zabs\ and \Zem\ to agree. The GRB progenitor is expected to originate in a low-metallicity region of star formation within the galaxy \citep{Woosley93, Metha+20}. \Zem\ and \Zabs, then, can be understood to trace the metallicity of two different subsets of the ISM of that galaxy. While \Zem\ measures the overall metallicity of star-forming gas, \Zabs\ must originate in a low-metallicity region of star-formation, and is therefore biased to prefer low-metallicities. For high-redshift galaxies with non-homogeneous metallicity distributions, these two metallicity measures are not expected to be equal. Therefore, by observing the difference between these two measured quantities in a representative sample of GRB hosts at redshift $z\sim 2-4$ and by comparing results to theoretical predictions from cosmological simulations, we have the opportunity to constrain subgrid models of metal mixing and enrichment.

We started exploring this idea in \citet{ZZ_TNG}, where predictions of the relationship between \Zabs\ and \Zem\ were constructed using the state-of-the-art cosmological magneto-hydrodynamic simulation IllustrisTNG. Motivated by approved JWST spectroscopic observations of a sample of GRB host galaxies with \Zabs\ already measured from the afterglow, in this paper, we extend our earlier analysis, constructing explicit predictions for the relationship between \Zabs\ and \Zem\ for two other large-volume hydrodynamic simulations: the original Illustris simulation \citep{Illustris1}, and the EAGLE simulation \citep{EAGLE}. Our aim is to identify which key features of the \Zabs\ versus \Zem\ relationship are shared across simulations, and which are unique and thus offer predictions that can be falsified by the upcoming JWST Cycle 1 data. 

This paper is organised as follows. In Section \ref{sec:simulations}, the three simulations analysed in our work are discussed, with a focus on the similarities and differences between the subgrid methods relevant for chemical enrichment. We outline our Monte-Carlo methodology for measuring the \Zabs-\Zem\ relation for these simulations in Section \ref{sec:methods}, and show the predicted form of this relationship for each simulation in Section \ref{sec:results}. In Section \ref{sec:comparisons}, we compare the predictions of the simulations, and calculate how many observational data-points would be necessary in order to rule out any of the models of galaxy evolution. We discuss our findings in Section \ref{sec:disco} and summarise our main results in Section \ref{sec:conc}.

\section{Simulations} \label{sec:simulations}

In this Section, we introduce the three cosmological simulations that we compare in this work. These are Illustris-1 from the Illustris simulation suite, Ref-L0100N1504 from the EAGLE simulation suite, and TNG100-1 from the IllustrisTNG simulation suite. Moving forward, we will refer to these simulations as Illustris, EAGLE, and TNG, for brevity. All three of these simulations have similar mass resolutions for baryonic particles and cover similar cosmic volumes of $\sim (100 $ cMpc$)^3$, but differ in many other respects. A broad overview of the similarities and differences between these three models is summarised in Table \ref{tab:sim_summary}.

In this Section, we will outline the most relevant details on the subgrid physical prescriptions of each model, focusing on the treatments of stellar and AGN feedback and metal enrichment. We will also explain which observables each simulation has been calibrated to, and whether there are any known tensions with observations. Readers who are already familiar with these three simulation suites or are not interested in the details of the sub-resolution models may wish to skip this section. Conversely, readers who wish to learn more about the wide variety of cosmological simulations used today are encouraged to read the review by \citet{Vogelsberger+20}. 

\begin{table*}
    \centering
    \begin{tabular}{p{5.2cm}|p{5.2cm}|p{5.2cm}}
    \textbf{EAGLE}& \textbf{Illustris} & \textbf{IllustrisTNG} \\
    Ref-L0100N1504 &Illustris-1 &  TNG100-1 \\
    \hline 
    (100 cMpc)$^3$ &(106.5 cMpc)$^3$&    (110.7 cMpc)$^3$ \\
    $m_{\textrm{baryon}} \sim 1.8 \times 10^6 M_\odot$ &$m_{\textrm{baryon}} \sim 1.3 \times 10^6 M_\odot$ &  $m_{\textrm{baryon}} \sim 1.4 \times 10^6 M_\odot$ \\
    Planck 2013 cosmology & WMAP-9 cosmology & Planck 2015 cosmology \\
    TreePM smooth particle hydrodynamics & AREPO moving mesh code  & AREPO moving mesh code \\
    \hline
    \citet{Chabrier03} IMF & \citet{Chabrier03} IMF & \citet{Chabrier03} IMF \\
    Stochastic star formation with $Z$-dependant density threshold & Stochastic star formation with fixed density threshold &  Stochastic star formation with fixed density threshold \\
    \citet{Portinari+98} stellar lifetimes & \citet{Portinari+98} stellar lifetimes & \citet{Portinari+98} stellar lifetimes \\
    Stochastic thermal stellar feedback &Kinetic stellar feedback &  Improved kinetic stellar feedback \\
    \hline
    \citet{Wiersma+09} cooling & \citet{Wiersma+09} cooling & \citet{Wiersma+09} cooling \\
    No self-shielding correction &Self-shielding correction &  Self-shielding correction \\
    11 elements tracked& 9 elements tracked & 9 elements + ``other metals" tracked \\
    Stellar yields of \citet{Portinari+98} and \citet{Marigo01} & Stellar yields of \citet{Portinari+98} and \citet{Karakas10} & Combined stellar yield model; see Table 2 of \citet{Pillepich+18}\\
    \hline
    BH seeds of mass $10^5 M_\odot / h$ & BH seeds of mass $10^5 M_\odot / h$ & BH seeds of mass $8 \times 10^5 M_\odot / h$ \\
    Bondi-Hoyle accretion, up to Eddington limit & Boosted Bondi-Hoyle accretion, up to Eddington limit & Bondi-Hoyle accretion, up to Eddington limit \\
    Stochastic thermal feedback for all AGN & Bubble model of \citet{Sijacki+07} for low accretion rate AGN & Stochastic kinetic feedback for low accretion rate AGN \\
    & Continuous thermal energy injection for rapidly accreting AGN & Continuous thermal energy injection for rapidly accreting AGN\\
    \hline
    No magnetism & No magnetism & Includes magnetism \\
    
    \end{tabular}
    \caption{Overview of the similarities and differences between the subgrid physical prescriptions that are most important in affecting the metallicity distribution of the ISM, for the three simulations compared in this work.}
    \label{tab:sim_summary}
\end{table*}

\subsection{EAGLE} \label{ssec:EAGLE}

The Evolution and Assembly of GaLaxies and their Environments project (EAGLE) is a large-scale hydrodynamical simulation suite that contains both dark matter and baryonic particles \citep{EAGLE}. Using the TreePM smoothed particle hydrodynamics code \textsc{Gadget 3} \citep{Springel05}, a large number (3.4 trillion in the simulation we are using) of dark-matter particles are allowed to evolve with time under the $\Lambda$CDM model of gravity, using the cosmological parameters reported in \citet{Planck13}. Unlike other large-scale simulations \citep[e.g.][]{Millennium, Uchuu}, EAGLE also simulates baryonic particles, allowing them to flow in the gravitational field created by the dark matter particles. The distribution of these baryonic particles then affect the distribution of dark matter in real-time, producing physically-motivated descriptions of galaxies without relying on semi-analytical modelling.

In EAGLE, star formation occurs when gas clouds cool and condense below a metallicity dependant threshold density \citep{Schaye04}. The metallicity-dependant cooling rates are taken from \citet{Wiersma+09}. As the simulation does not have the resolution to model individual stars, simple stellar populations of mass $\sim 10^6 M_\odot$ are formed, with metallicities inherited from the gas cells they are formed from, and masses taken from a \citet{Chabrier03} initial mass function (IMF). These stellar populations are then evolved following the models of \citet{Portinari+98}, releasing mass and metals throughout their lives as the stars they are composed of collapse. The metallicity yields of stars that end their lives as core collapse supernovae are taken from \citet{Portinari+98} and \citet{Marigo01}.

Historically, feedback from supernovae and AGN has been difficult to model in hydrodynamical simulations. Thermally injected feedback radiates away too efficiently to drive shocks, a challenge termed the `overcooling' problem \citep[e.g.][]{Katz+96, Springel+Hernquist03, Naab+Ostriker17}. EAGLE solves this problem by increasing the amount of energy injected by core collapse supernovae such that any cell receiving energy from supernova feedback must undergo a temperature change of at least $\Delta T = 10^{7.5}$ K, while simultaneously decreasing the frequency of feedback events so that the total amount of energy injected from supernovae is conserved. The higher temperature of the heated gas particles increases the time taken for thermal energy to be dissipated, allowing stellar feedback to be more efficient at driving galactic winds \citep{Dalla-Vecchia+Schaye12}. This approach, known as \textit{stochastic thermal feedback}, leads to burstier feedback with more realistic thermal gradients, and it is used to model feedback from both supernovae and AGN.%, both in the low-and high-accretion rate regimes

EAGLE was calibrated to reproduce three key observables at redshift zero: the galaxy stellar mass function (GSMF); the galaxy mass-size relation; and the relationship between the stellar mass of galaxies and the masses of their central supermassive black holes. While it has been found to further reproduce many other galaxy-scale properties of the local universe, some tensions have been found. For example, \citet{Yates+21} suggested that simulation Ref-L0100N1504 (the reference run we are using) may retain too many metals in the ISM compared to observations, in turn affecting the mass-metallicity relation (MZR) at redshift zero \citep{EAGLE}. Also, the specific star-formation rate (sSFR) at redshift zero for low-mass galaxies is lower than observed (see Figures 11 and 13 of \citealt{EAGLE}), both of which may significantly impact the population of GRB host galaxies found within this simulation. Finally, we note that the number density of large galaxies produced at $z \sim 2$ in the EAGLE simulation is substantially lower than the number produced by the Illustris/TNG simulations \citep{Oppenheimer+20}.

\subsection{Illustris} \label{ssec:Illustris}

Similarly to EAGLE, the Illustris project \citep{Illustris1, Illustris2, Illustris3, Illustris4} also seeks so simulate large-scale structure formation in a $\Lambda$CDM universe (although with a slightly different set of cosmological parameters from WMAP-9; \citealt{WMAP9}), accounting for baryonic physics using subresolution models in order to create theoretical predictions of galaxy properties on a cosmic scale.

Unlike EAGLE, which uses smoothed-particle hydrodynamics (SPH), the Illustris simulation suite takes advantage of the moving-mesh code \textsc{arepo} \citep{Springel10}. By allowing a set of mesh-generating points to flow freely under a velocity field, a 3D Voronoi tessellation is constructed at each timestep. Mass, metals, momentum, and energy are then allowed to flow between these Voronoi cells using a mesh-based solution to the laws of ideal hydrodynamics, which in turn sets the velocities of the mesh-generating points for the next timestep. This mathematical approach combines aspects of SPH with Eulerian mesh-based methods, simultaneously avoiding numerical issues that both other methods possess \citep{Vogelsberger+13}.

Star formation occurs stochastically in gas cells with a density greater than $0.1$ cm$^{-3}$ (unlike EAGLE, this threshold is not metallicity-dependant). Metal-line cooling is computed for each cell based on the cooling rates of \citet{Wiersma+09}, but with an additional factor for self-shielding based on the results of \citet{Rahmati+13}. Stellar evolution is modelled to follow the evolutionary tracks of \citet{Portinari+98}, with metal yields informed by \citet{Portinari+98} and \citet{Karakas10}. 

To overcome the overcooling problem, in Illustris, stellar feedback is injected kinetically rather than thermally. Star forming gas cells stochastically transform a fraction of their mass into `wind particles', which are launched in a bipolar direction out of the galactic plane. The metallicity of these wind particles is set to be $0.4\times $the metallicity of the gas cell from which they are launched, in order to help match the MZR for low-mass galaxies \citep{Zahid+12}.
While they are travelling, these wind particles are not permitted to interact with other particles in the simulation, except gravitationally.
The velocity of these wind particles are set to be proportional to the velocity dispersion of the dark matter halo, and their masses are then computed based on the available energy from all core collapse supernovae.
Once the wind particles enters a low density region (below $5\%$ of the density required for star formation), they dissolve, depositing all of their mass, metals, momentum, and thermal energy into the closest gas cell.

This prescription neglects delay-times associated with core-collapse supernovae, which are short. Furthermore, the process of decoupling wind particles from the ISM may effect galaxy properties on large scales
\citet{Dalla-Vecchia+Schaye08}, 
and will undoubtedly impact the small-scale metallicity structure of the ISM.

Three forms of AGN feedback are accounted for: thermal, mechanical, and electromagnetic, to reflect the multitude of distinct physical processes that contribute to AGN feedback \citep[e.g.][]{Begelman14}. For black holes with accretion rates below $5\%$ of the Eddington limit, mechanical energy is injected directly into a spherical bubble at a distance of $\lesssim 100$ kpc from the AGN, following the phenomenological model of \citet{Sijacki+07}. For black holes with larger accretion rates, thermal energy is continuously injected into the environment around the AGN, to simulate quasar activity. All accreting black holes are also presumed to produce ionising radiation that heats up the gas cells surrounding them, as an additional form of feedback.

The Illustris simulation suite was calibrated to reproduce the redshift zero GSMF, the stellar-to-halo mass relation ($M_*-M_{\textrm halo}$) at $z = 0$, and the star formation rate density as a function of cosmic time. Similarly to EAGLE, Illustris has shown broad agreement to a range of galaxy-scale observations at low and intermediate redshifts. However, \citet{Nelson+15} list several tensions between this simulation and observational data. Relevant to our work, the Illustris100-1 simulation has been found to overproduce both large ($M_* \gtrsim 10^{11.5}M_\odot$) and small ($M_* \lesssim 10^{10}M_\odot$) galaxies. Furthermore, in that study, galaxies of mass $M_* \lesssim 10^{10.7}M_\odot$ were found to be too spatially extended, and the gas fraction of galaxy groups to be too low.

\subsection{IllustrisTNG} \label{ssec:TNG}

To resolve these and other tensions, and to take advantage of newly discovered physical models and improved numerical recipes, The Next Generation Illustris Simulation (IllustrisTNG) was constructed as a sequel to the Illustris simulation suite \citep{TNG1, TNG2, TNG3, TNG4, TNG5}. Both Illustris and IllustrisTNG are run using the \textsc{arepo} moving mesh code, and the TNG100-1 simulation has a comparable volume and resolution to the Illustris-1 simulation. 

One major change from Illustris to IllustrisTNG is the incorporation of magnetic fields. The inclusion of this physics significantly increases the accretion rates of black holes and the gas fractions of small halos ($M < 10^{12} M_\odot$), and significantly suppresses star formation at $z \lesssim 3$, especially in large halos \citep{Pillepich+18}.

The TNG simulations use the same overall physical prescriptions for star formation and stellar feedback, but with a few important changes. Stellar yield tables are constructed from a variety of sources \citep{Nomoto+97, Portinari+98, Kobayashi+06, Karakas10, Doherty+14, Fishlock+14}, covering a range of initial masses and metallicities. Stellar winds are still launched from star-forming gas cells, but (1) winds are launched isotropically rather than preferentially out of the galactic plane; (2) a wind velocity floor is imposed to ensure that the mass of wind particles is not too high; (3) winds are faster and are dependant on redshift; (4) winds from low-metallicity gas cells are taken to be more energetic; and (5) ten percent of wind energy is injected as thermal energy, rather than kinetic energy, to prevent shocks from driving spurious star formation. Overall, the winds in TNG are faster than those in Illustris, and more efficient at preventing star formation in all galaxies, regardless of their redshift or mass.

The low-accretion mode of black-hole feedback has also been updated. In TNG, momentum is added in a random direction once a certain energy threshold is reached -- similar to the prescription used in the EAGLE subgrid model, but delivering energy kinetically, rather than thermally. Both other methods of black-hole feedback, thermal and electromagnetic, remain qualitatively the same as in Illustris.

The TNG simulations are calibrated to reproduce six observables of galaxy populations: (1) the cosmic star formation density as a function of redshift; (2) the GSMF at $z=0$; (3) the $M_*-M_{\textrm halo}$ relation at $z=0$; (4) the $M_{\textrm BH}-M_{\textrm halo}$ relation at $z=0$; (5) the gas fraction of galaxies; and (6) the sizes of galaxies. The stellar feedback models implemented in IllustrisTNG show improved agreements with observations over the Illustris model for all six of these observables. 
However, % some notable tensions with observations have been reported.
\citet{Yates+21} note that the star formation rate in TNG peaks at $z\sim 3$, in tension with observational determinations of the cosmic star formation history that infer a $z\sim 2$ peak location (e.g. see \citealt{Madau+Dickinson14} and \citealt{Driver+18}). This may lead to an excess of metals being produced at lower redshifts compared to other simulations.

\section{Methods} \label{sec:methods}

For each simulation, the predicted difference between \Zabs\ and \Zem\ for GRB host galaxies was determined at four redshifts: $z=1.74, z=2.32, z=3.0$, and $z=4.0$. The four snapshots closest to these redshifts were downloaded for each simulation.\footnote{For Illustris, the snapshot numbers were 54, 60, 66, and 71, all available at \url{http://www.illustris-project.org/data/}. For TNG, snapshots 21, 25, 30, and 36 were downloaded from \url{http://www.tng-project.org/data/}. For EAGLE, snapshots 10, 12, 14 and 16 were downloaded from \url{http://virgodb.dur.ac.uk/}.}

In the three simulations, there exists a known issue with the \textsc{subfind} algorithm occasionally erroneously identifying overdensities in subhalos as satellite galaxies \citet{Pillepich+18}. To circumvent this problem, we follow the prescription for identifying galaxies described in \citet{Metha+20}, whereby a galaxy is defined as a collection of star-forming baryonic particles that could not be visually separated by typical observations. In practise, this is done by merging all gas particles identified with subhalos whose half-star radii overlap in 3D. This definition ensures that galaxies in the simulations match what is observed in large, high-redshift surveys.
%For the EAGLE simulation, this step is not necessary and we use the results from the subfind algorithm in the publicly released data. 
% @@ Reworded a bit above to address your concerns
%taken, because (i) the issue with the subfind algorithm has not been reported \textbf{that I know of}, and (ii) the data structure of the EAGLE simulation makes this kind of analysis much more difficult \textbf{but can I say this in a scientific journal? Do I have to be more specific?}. 

Following \citet{Metha+20}, the probability of each gas cell hosting a GRB was determined by multiplying the star-formation rate of each gas cell by a metallicity bias function. This model takes advantage of the short delay time distribution for long gamma-ray bursts, which implies that the rate of GRBs should be approximately proportional to the birth rate of GRB progenitor stars. 

\begin{equation}
    \label{eq:pr_host_cell}
    Pr^{(GRB)} = \frac{\rho^{\text(GRB)}_\text{cell}}{ \sum_{\text{all gas cells}}\rho^{\text(GRB)}_\text{cell}}.
\end{equation}

As in \citet{ZZ_TNG}, the metallicity bias function for GRB progenitor stars is taken to be a simple step function:

\begin{equation} \label{eq:cutoff_bias}
    \kappa(Z)=\begin{cases}
      \kappa_{0}, & \text{if}\ Z < Z_{\rm max} \\
      0, & \text{otherwise}.
    \end{cases}
\end{equation}

Five different cutoff models were explored, with \Zmax$=0.2Z_\odot, 0.4Z_\odot, 0.6Z_\odot, 0.8Z_\odot,$ and $1.0Z_\odot$.
We also tested one model with no metallicity bias where the rate of GRB formation was assumed to be directly proportional to the star formation rate, in order to separate the effects of a metallicity bias from geometric effects. (In \citet{ZZ_TNG}, even a model with no metallicity bias was found to be able to impart a significant difference between \Zabs\ and \Zem.)

For each snapshot of each simulation, under each GRB metallicity bias model, 2000 gas cells were selected with random line-of-sight directions from the simulation directly, with the likelihood of each gas cell being selected weighted by its probability to form a GRB progenitor (Equation \ref{eq:pr_host_cell}). This prescription automatically selects GRB host galaxies from the simulation volumes in a consistent way, allowing one host galaxy to host multiple GRBs with their jets oriented in different directions.\footnote{We note that this prescription is subtly different from the methodology used in \citet{ZZ_TNG}, wherein one GRB host gas cell is selected for each galaxy before a sample of GRB host galaxies are selected.}

We take our definitions for \Zem\ and \Zabs\ from \citet{Metha+20}. Because emission line ratios depend on flux from star-forming \Hii regions, we define \Zem\ to be the star-formation rate weighted average metallicity of all gas cells in a galaxy:

\begin{equation} \label{eq:z_emiss}
    Z_{\rm emiss} := \frac{\sum_{\text{all gas cells}} Z_{\rm cell} \times {\rm SFR}_{\rm cell}}{\sum_{\text{all gas cells}}{\rm SFR}_{\rm cell}}.
\end{equation}

On the other hand, \Zabs\ is not biased by star-formation. Rather, it represents the mass fraction of metals along a line-of-sight (LOS) from the source to the observer:

\begin{equation}
    \label{eq:Z_abs}
     Z_{\rm abs} := \frac{\sum_{\text{gas cells along LOS}} Z_{\rm cell} \times \rho_{\rm cell} \times l_{\rm cell}}{\sum_{\text{gas cells along LOS}} \rho_{\rm cell} \times l_{\rm cell}}.
\end{equation}

Unlike Illustris and TNG, the EAGLE simulation is not a mesh-based cosmological hydrodynamic simulation. To make this definition compatible with the smoothed particle hydrodynamics scheme of EAGLE, we instead treat the metallicity and density of the ISM at any point to be equal to the metallicity and density of the closest gas particle. This is mathematically equivalent to constructing a 3D Voronoi tessellation for the EAGLE simulation using the gas particles as mesh-generating points, but is much less computationally intense.

\section{Results} \label{sec:results}

\subsection{Predictions from TNG} \label{ssec:TNG_ZZ_rel}

\begin{figure*}
	\includegraphics[width=0.49\textwidth]{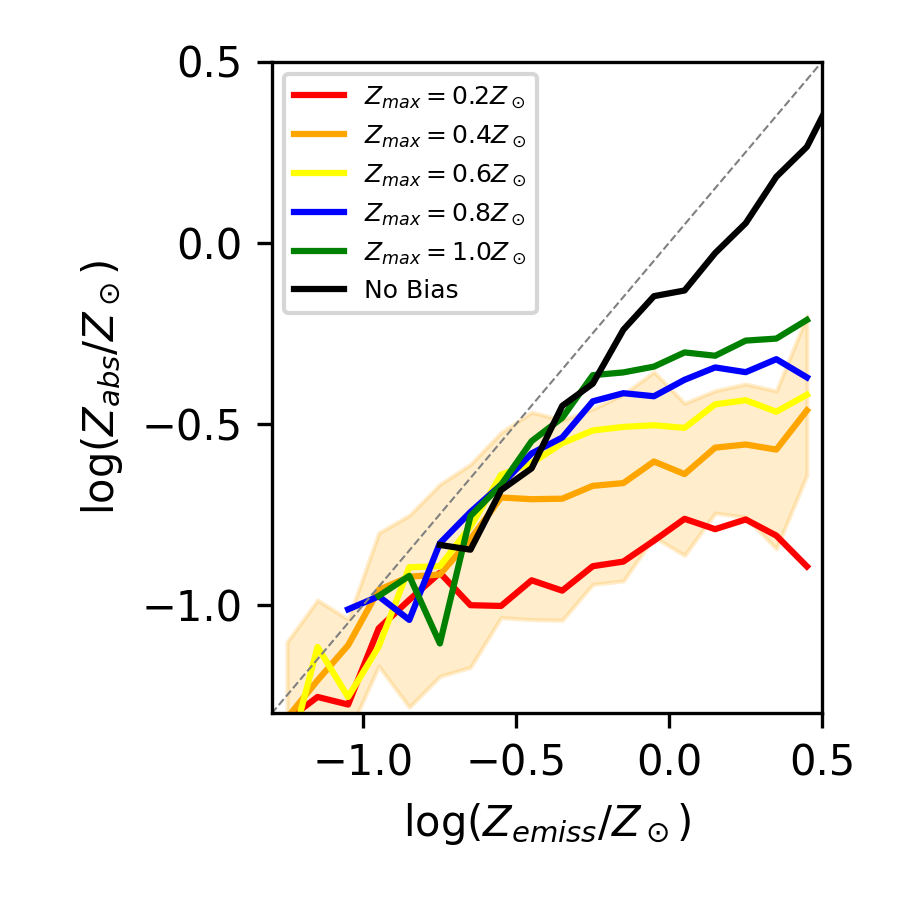}
	\includegraphics[width=0.49\textwidth]{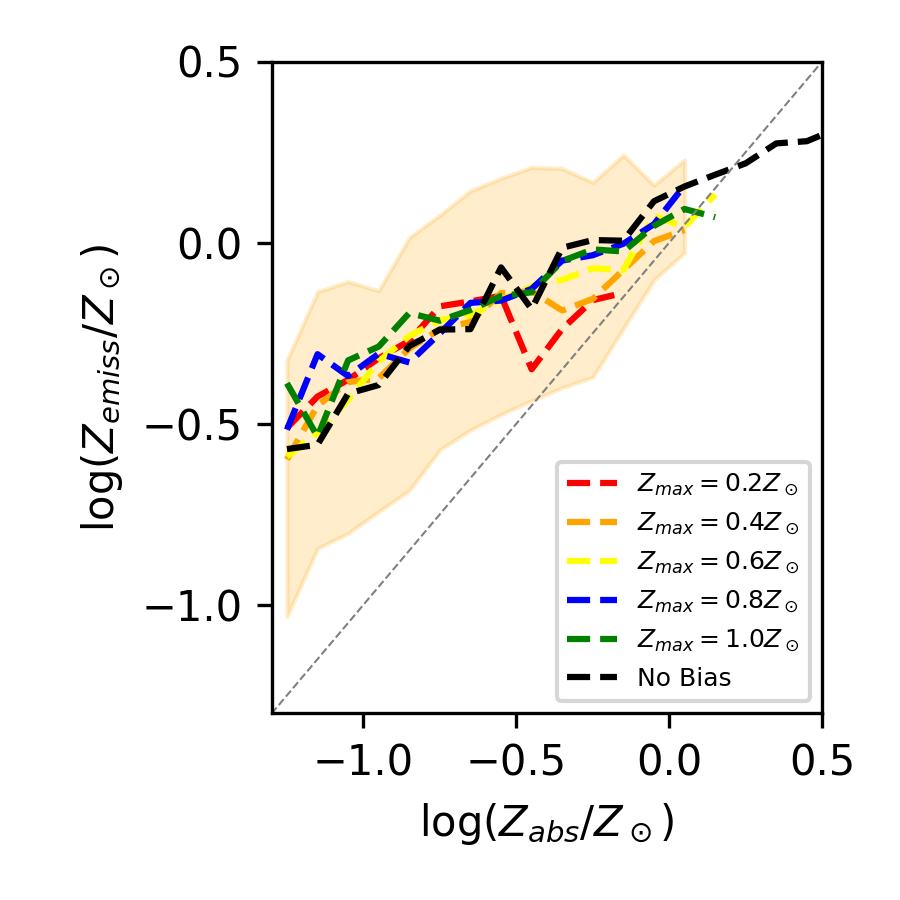}
    \caption{\emph{Left:} The median value of $Z_{\rm abs}$ for TNG GRB host galaxies with a fixed value of $Z_{\rm emiss}$. \emph{Right:} The median value of $Z_{\rm emiss}$ for TNG GRB host galaxies with a fixed value of $Z_{\rm abs}$. Both of these plots show the distribution at a redshift of $z=2.32$. An error region showing the $16$th- and $84$th-percentiles for a GRB metallicity bias function with $Z_{\rm max}=0.4Z_\odot$ is shown in both panels - error regions are of a similar size for all othe models.}
    \label{fig:TNG_z=2.33_median}
\end{figure*}

As a reference starting point for our results, in Figure \ref{fig:TNG_z=2.33_median}, we show the median values for \Zabs\ at a fixed value of \Zem\ (left), and \Zem\ at a fixed value of \Zabs\ (right), for the 2000 GRBs simulated at snapshot 30 of the TNG simulation used at a redshift of $z=2.32$, investigating the effects of changing the cutoff metallicity for GRB formation in this simulation. Similar plots are also shown in \citet{ZZ_TNG}, but for the convenience of the reader we re-present them here along with our main relevant conclusions of that paper.

From these figures, we see that the shape of the median \Zabs-\Zem\ relation looks different depending on which axis is treated as the independent variable. For a given value of \Zem, the median value of \Zabs\ roughly follows \Zem\ when \Zem$\lesssim$\Zmax, the maximum metallicity for GRB formation. After this value, the median value of \Zabs\ grows much slower. Such a feature is to be expected from the nature of our model, as in galaxies with a higher average metallicity, the location of GRB formation will be restricted to the lower-metallicity regions, and lines of sight originating from these locations are less likely to travel through high-metallicity regions of star formation.\footnote{This is true regardless of whether the metallicity gradient of the galaxy is negative (as is seen in galaxies in the local universe), positive (as is found for high-redshift galaxies in the TNG simulation by \citealt{Hemler+21}), or flat, as long as the galaxy does not have a uniform metallicity throughout. We discuss this point further in Section \ref{sec:disco}.} 

When \Zabs\ is fixed, the median value of \Zem\ shows little dependence on the threshold metallicity for GRB formation \Zmax. Instead, \Zem\ evolves linearly with \Zabs, with a slope of $\sim 0.4$ \citep{Metha+20}. The shape of this trend depends on the internal metallicity structures of the population of star-forming galaxies at this redshift, and is different for different simulations (see Figures \ref{fig:EAG_z=2.33_median} and \ref{fig:OG_z=2.33_median}, and discussion in Section \ref{sec:comparisons}). 

\begin{figure*}
	\includegraphics[width=0.49\textwidth]{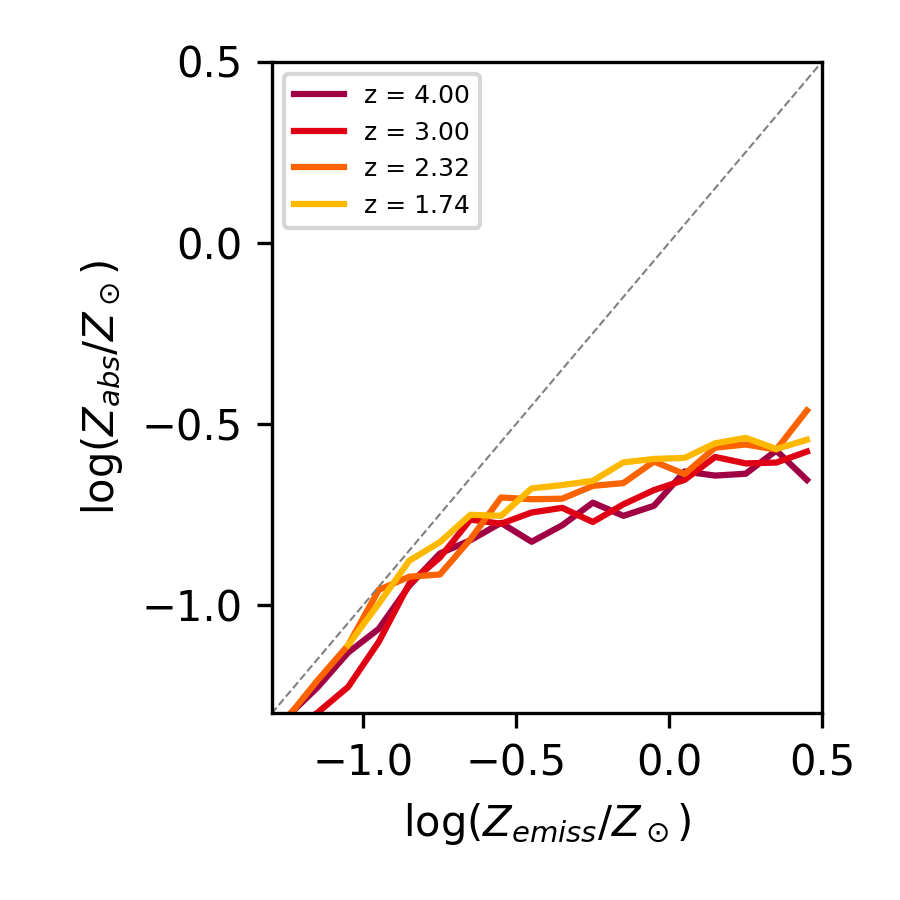}
	\includegraphics[width=0.49\textwidth]{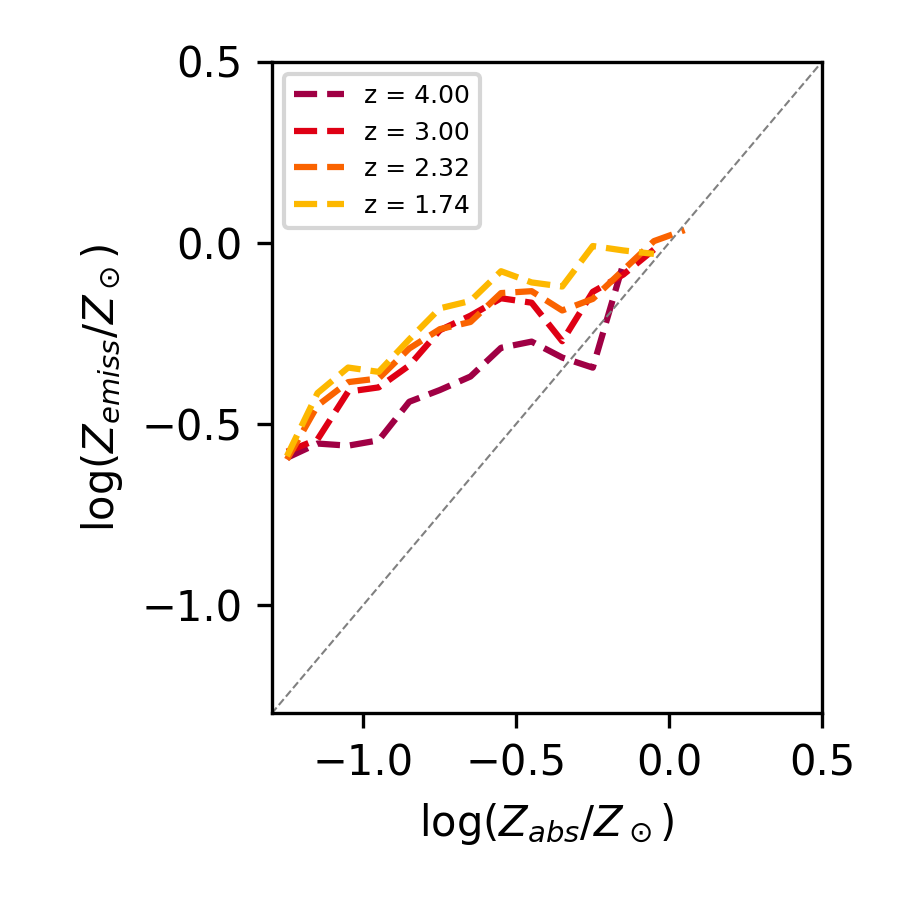}
    \caption{\emph{Left:} Redshift evolution of the median value of $Z_{\rm abs}$ for TNG galaxies with a fixed value of $Z_{\rm emiss}$. \emph{Right:} Redshift evolution of the median value of $Z_{\rm emiss}$ for TNG galaxies with a fixed value of $Z_{\rm abs}$. All of these plots assume a maximum metallicity for GRB progenitors of $Z_{\rm max}=0.4Z_\odot$.}
    \label{fig:TNG_z-dep}
\end{figure*}

In Figure \ref{fig:TNG_z-dep}, we show how these relationships change with redshift, using a value of $Z_{\rm max}=0.4Z_\odot$ as our fiducial model. We find that the median value of \Zabs\ as a function of \Zem\ changes very slowly, with higher-redshift galaxies harbouring lower-metallicity regions of star-forming gas. This relationship is more clearly observed in the right hand panel, where the median value of \Zem\ for given values of \Zabs\ is found to drop significantly from $z=1.74$ to $z=4.0$. This feature reflects the cosmic metal enrichment taking place between these two epochs in the TNG simulation. The same trends are seen for all cutoffs explored in this section, implying that the redshift evolution of the \Zabs-\Zem\ relation is mild for all values of \Zmax.

\subsection{Predictions from EAGLE} \label{ssec:EAGLE_ZZ_rel}

\begin{figure*}
	\includegraphics[width=0.49\textwidth]{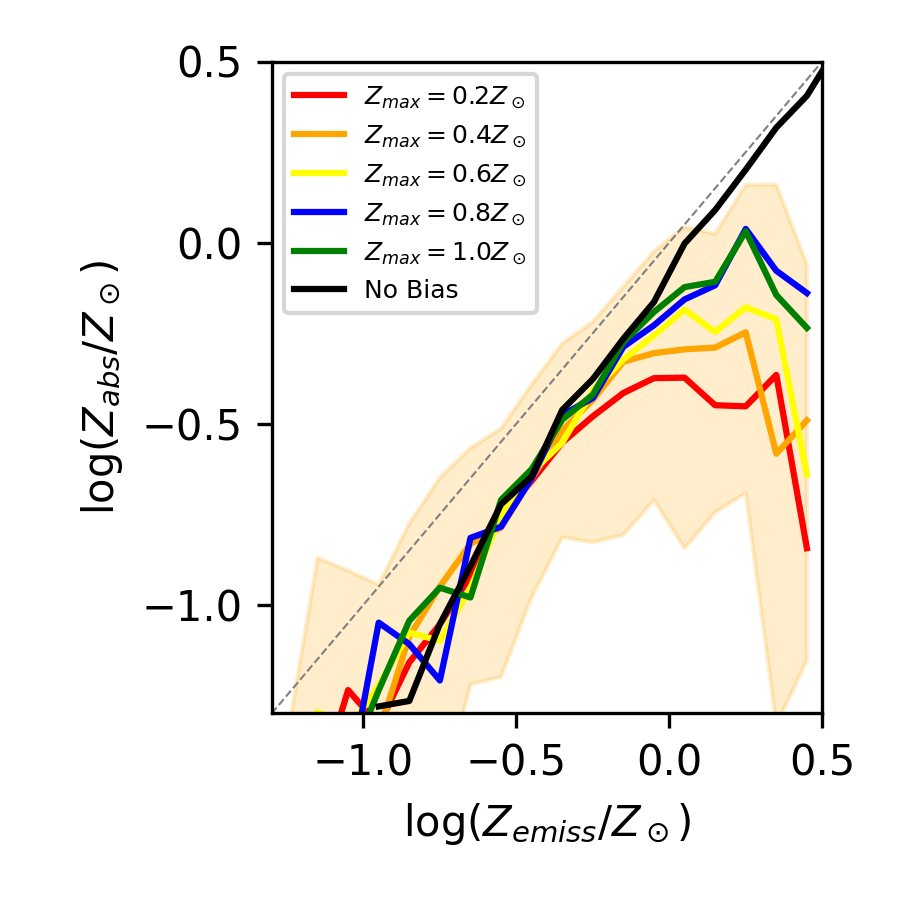}
	\includegraphics[width=0.49\textwidth]{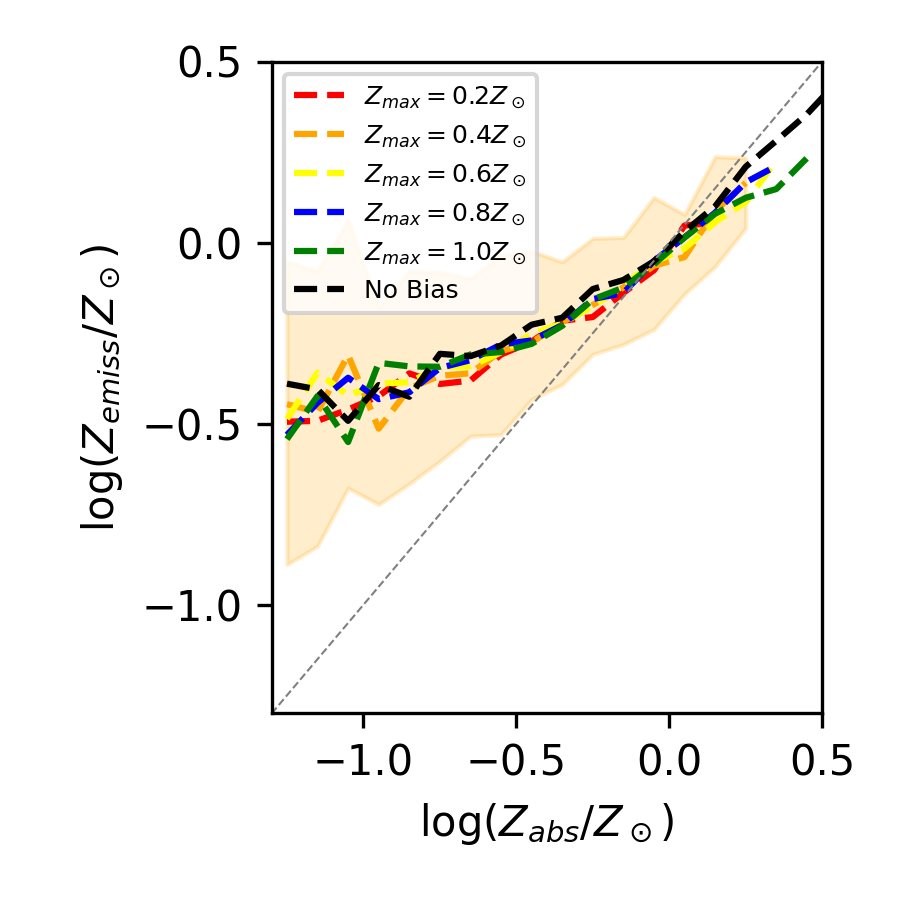}
    \caption{\emph{Left:} The median value of $Z_{\rm abs}$ for EAGLE galaxies with a fixed value of $Z_{\rm emiss}$. \emph{Right:} The median value of $Z_{\rm emiss}$ for EAGLE galaxies with a fixed value of $Z_{\rm abs}$. Both of these plots show the distribution at a redshift of $z=2.32$. An error region showing the $16$th- and $84$th-percentiles for a GRB metallicity bias function with $Z_{\rm max}=0.4Z_\odot$ is shown in both panels - error regions are of a similar size for all models.}
    \label{fig:EAG_z=2.33_median}
\end{figure*}

In Figure \ref{fig:EAG_z=2.33_median}, we plot an analog of Figure \ref{fig:TNG_z=2.33_median} using data from the EAGLE RefL0100N1504 simulation at a redshift of 2.33.

\begin{figure*}
	\includegraphics[width=0.49\textwidth]{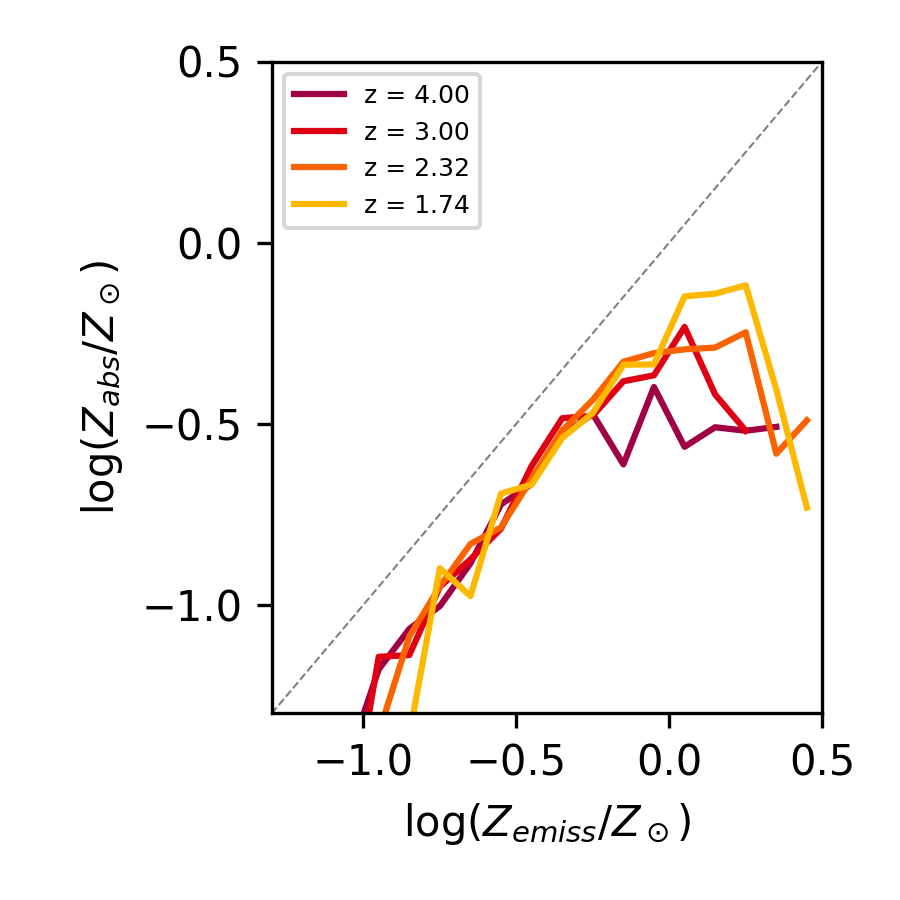}
	\includegraphics[width=0.49\textwidth]{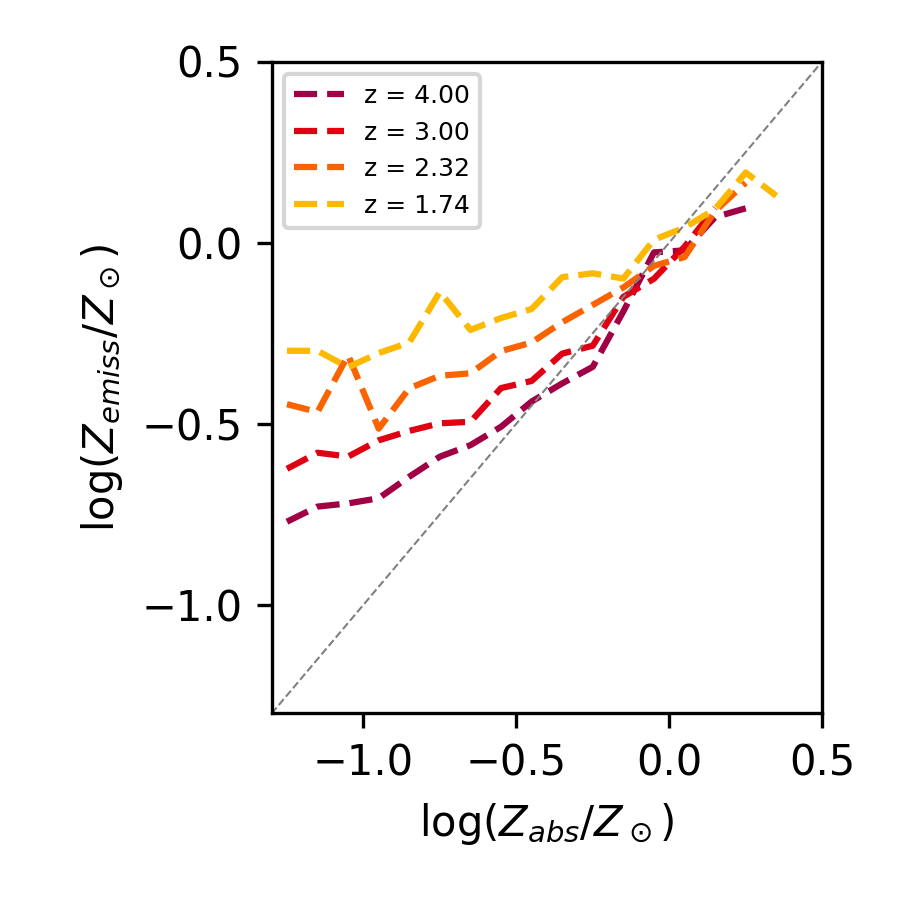}
    \caption{\emph{Left:} Redshift evolution of the median value of $Z_{\rm abs}$ for EAGLE galaxies with a fixed value of $Z_{\rm emiss}$. \emph{Right:}  Redshift evolution of the median value of $Z_{\rm emiss}$ for EAGLE galaxies with a fixed value of $Z_{\rm abs}$. All of these plots assume a maximum metallicity for GRB progenitors of $Z_{\rm max}=0.4Z_\odot$.}
    \label{fig:EAGLE_z-dep}
\end{figure*}

Interestingly, we see that when \Zem$\lesssim 0.5Z_\odot$, the median value of \Zabs\ at fixed \Zem\ does not seem to depend on the cutoff metallicity for GRB formation. This feature was seen at all redshifts in the EAGLE snapshots investigated in this work. We can explain this feature by noting that (i) \Zabs\ and \Zem\ are expected to be equal when a galaxy is chemically homogeneous, and (ii) low metallicity galaxies tend to have lower masses. Therefore, such low metallicity galaxies are more likely to be well-mixed due to efficient feedback processes being able to operate on the scales of these dwarf galaxies.

This result is consistent with, and extends on, the findings of \citet{Tissera+22}, who show that galaxies produced in the EAGLE simulation tend to have metallicity gradients close to zero, with little dependence on stellar mass or redshift. Such a zero metallicity gradient could be consistent with one of two scenarios: either the galaxy is chemically well-mixed; or it contains asymmetric metallicity fluctuations that cannot be captured by a simple linear gradient model. Our \Zabs-\Zem\ relation shows that both of these possibilities occur in the EAGLE simulation: larger galaxies tend to have a non-uniform, non-radial metallicity structure, whereas smaller galaxies tend to be more chemically homogeneous. \citet{Tissera+22} also show that low- and intermediate-mass galaxies in EAGLE have an enhanced star formation efficiency, which would also lead to an increased amount of feedback, giving a physical explanation to the higher degree of homogeneity seen in these smaller galaxies.

When no metallicity bias is imposed, for low-metallicity galaxies, the value of \Zabs\ is substantially lower than the reported value for \Zem. An investigation of the data revealed that while the bulk of the population follow a trend where \Zabs$\approx$\Zem, a small population of galaxies with extremely low values of \Zabs\ ($10^{-5}-10^{-2}Z_\odot$) was also found. These anomalously low galaxies existed at all values of \Zem, but began to dominate the population at low metallicities, where the number of regular galaxies was lower. We discuss this population further in Section \ref{ssec:low_Zabs_hosts}.

The trend of the median value of \Zem\ at any given \Zabs\ is also seen in TNG. However, the shapes of these relations are very different between the two simulations. While the relationship for TNG was nearly linear, the relationship seen for EAGLE appears to be closer to a quadratic, with a flatter slope for \Zabs$\lesssim 0.3Z_\odot$ and a steeper slope elsewhere. This indicates that the predictions of the internal structure of the ISM generally differs between simulations that use different physical prescriptions for metal production and transport.

In Figure \ref{fig:EAGLE_z-dep} we show how these relationships depend on redshift for the model with $Z_{\rm max}=0.4Z_\odot$. We see that while the value of \Zabs\ at fixed \Zem\ shows very little redshift evolution (left panel), the median value of \Zem\ at fixed \Zabs\ is much higher for lower-redshift galaxies. This effect is especially significant at low values of \Zabs. Such an evolution reflects chemical enrichment in the EAGLE simulation over cosmic time, and is consistent with what is observed in the TNG simulation.

\subsubsection{The population of extremely low metallicity GRB absorbing systems in EAGLE} \label{ssec:low_Zabs_hosts}

In every snapshot of the EAGLE simulation we considered, for every value of \Zmax\ tested, a population of galaxies with \logZabs$<-2$ was identified. This is more than an order of magnitude lower than the median value of \Zabs\ found in each galaxy sample. This population is most common at high redshift where the metallicity of galaxies is lower, and for low values of \Zmax, where extremely low values of \Zabs\ are favoured. At $z=4$ with a cutoff of \Zmax$=0.2Z_\odot$, this low metallicity sample makes up $14.8\%$ of GRB hosts seen in the EAGLE simulation. At lower redshifts $(z \leq 2.3)$, for all values of \Zmax\ tested, this low metallicity population made up less than $5.5\%$ of GRB hosts. The median stellar masses of these galaxies are generally low ($M_* \approx 1-6 \times 10^6 \odot$, with lower median masses at higher redshifts), representing typically dwarf galaxies with only a few stellar population particles, but exhibiting a wide range, from galaxies with no stars (caught on the cusp of star formation), to large Milky-way size discs with $M_* \sim 10^{11} M_\odot$. Such an extreme low metallicity sub-population is not seen in our analysis of either the Illustris or the TNG simulations. 

Further analysis of this population reveals that the values of \Zabs\ measured for these galaxies appear to be not inconsistent with the metallicities of the gas particles from which the GRB originated. Such extremely low metallicity systems are associated instead to extremely low metallicity star-forming gas particles, with lines-of-sight directed away from the more enriched regions of these galaxies, which contribute to a higher \Zem.

The existence of such systems implies that galaxies in the EAGLE simulation contain pockets of star-forming gas with extremely low metallicities. In turn, this suggests that galactic fountains are inefficient at mixing metals into the outer regions of galaxies at redshifts of $z \sim 1.8-4$ in the EAGLE simulation. This may be due to the lack of metal diffusion between SPH particles in the EAGLE simulation, as no metal diffusion between SPH particles was implemented in EAGLE due to the high uncertainties on mixing coefficients \citep{EAGLE}. This allows extremely low-metallicity regions of gas to persist for longer in EAGLE than they would in Illustris or TNG, where metals are allowed to flow between different Voronoi cells. A second possible cause for these low-metallicity regions could be attributed to lower efficiency of the feedback system implemented in EAGLE for low-mass galaxies, as stochastic thermal feedback is less likely to occur in systems with a low number of star particles. 

Whether these systems are artefacts of the numerical choices made when building EAGLE or represent a true population of high-redshift galaxies with inefficient galactic fountains remains to be seen. Presently, only two GRB host galaxies with \logZabs$<-2$ have been reported: the host of GRB050730A at a redshift of $z=3.97$ with \logZabs$= - 2.31 \pm 0.1$ \citep{Wiseman+17}; and the host of GRB140311A at a redshift of $z=4.96$ with \logZabs$=-2.00 \pm 0.11$ \citep{Bolmer+19}. 
We describe how measurements of \Zem\ and \Zabs\ for the same GRB host galaxy can help confirm or falsify the presence of the poorly-mixed GRB host systems predicted by EAGLE in Section \ref{sec:n_to_constrain_models}.

\subsection{Predictions from Illustris} \label{ssec:Illustris_ZZ_rel}

\begin{figure*}
	\includegraphics[width=0.49\textwidth]{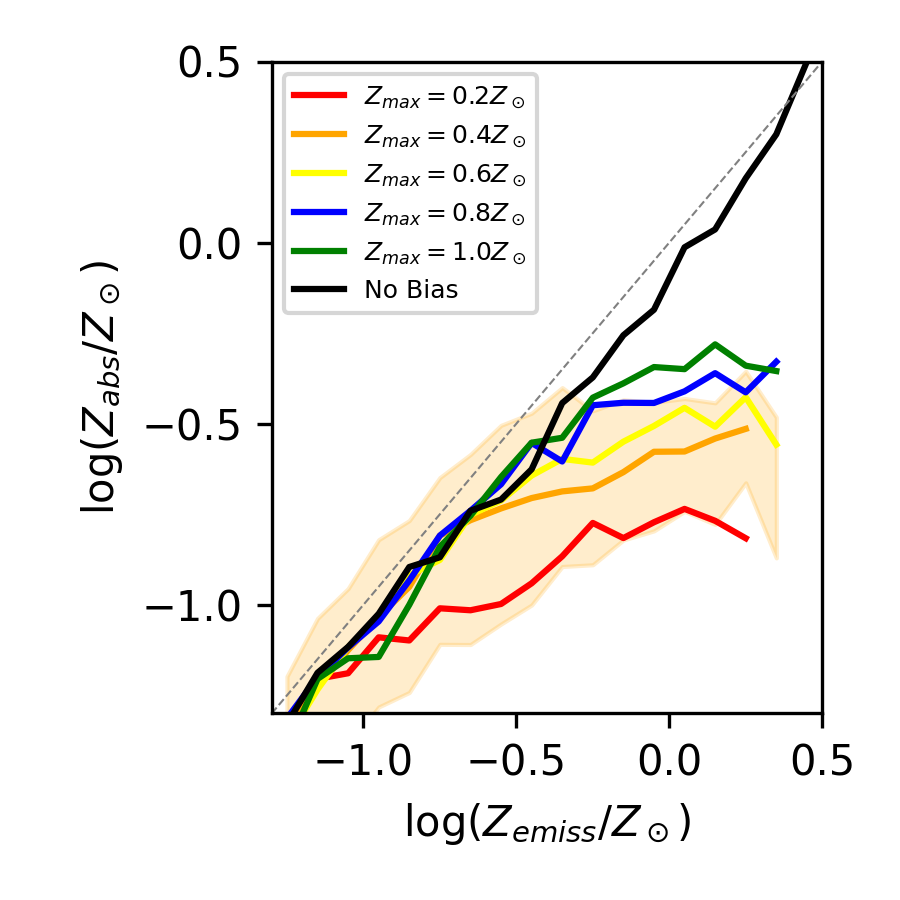}
	\includegraphics[width=0.49\textwidth]{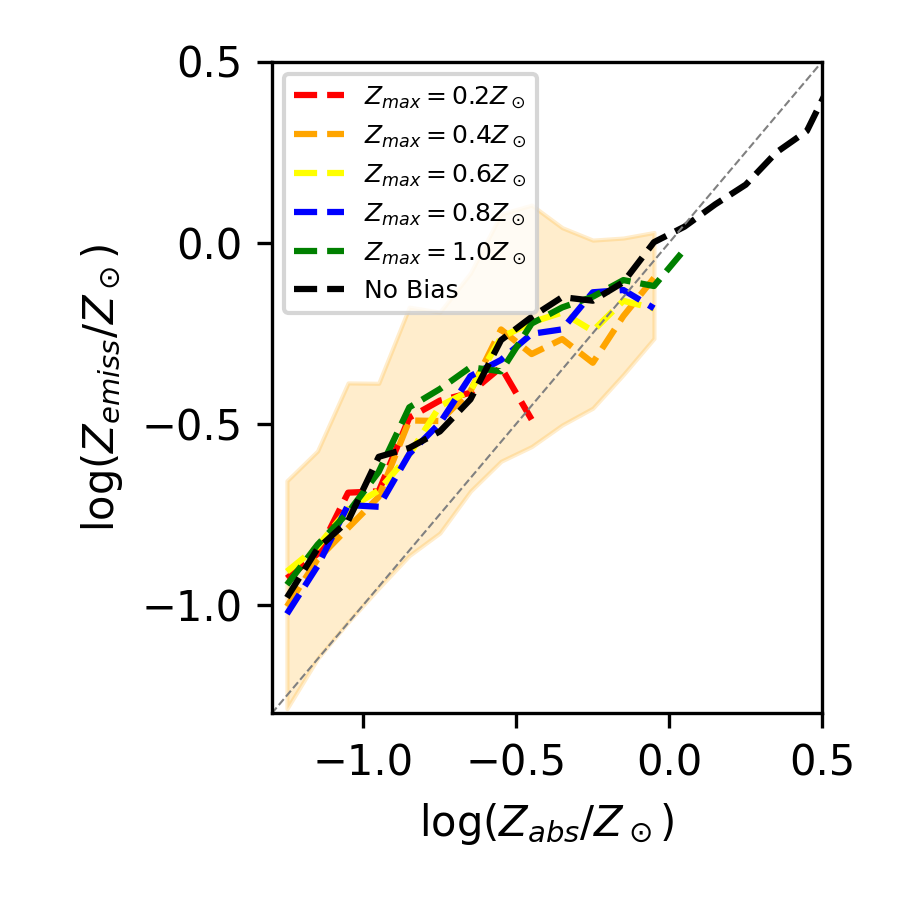}
    \caption{\emph{Left:} The median value of $Z_{\rm abs}$ for Illustris galaxies with a fixed value of $Z_{\rm emiss}$. \emph{Right:} The median value of $Z_{\rm emiss}$ for Illustris galaxies with a fixed value of $Z_{\rm abs}$. Both of these plots show the distribution at a redshift of $z=2.32$. An error region showing the $16$th- and $84$th-percentiles for a GRB metallicity bias function with $Z_{\rm max}=0.4Z_\odot$ is shown in both panels - error regions are of a similar size for all models.}
    \label{fig:OG_z=2.33_median}
\end{figure*}

In Figure \ref{fig:OG_z=2.33_median}, we show an analogous figure to Figure \ref{fig:TNG_z=2.33_median} for the sample of 2000 GRB host galaxies drawn from the Illustris simulation at $z=2.32$.

In the left hand panel, the median value of \Zabs\ at a fixed value of \Zem\ is shown. Unlike EAGLE, this figure shows much greater agreement to the \Zabs-\Zem\ relation predicted using the TNG simulation, with the predictions for a model with a cutoff of \Zmax\ only agreeing with the metallicity bias-free model when \Zem$\leq$\Zmax.

In the right hand panel, the median value of \Zabs\ at fixed \Zem\ is displayed. Unlike EAGLE, and similarly to TNG, this relationship appears to be linear. However, the slope of this relation appears to be very close to 1. As this holds for all values of \Zmax\ tested, we ascribe this relation to be driven more by the population of galaxies produced by Illustris rather than any details on the model of GRB formation. 

\begin{figure*}
	\includegraphics[width=0.49\textwidth]{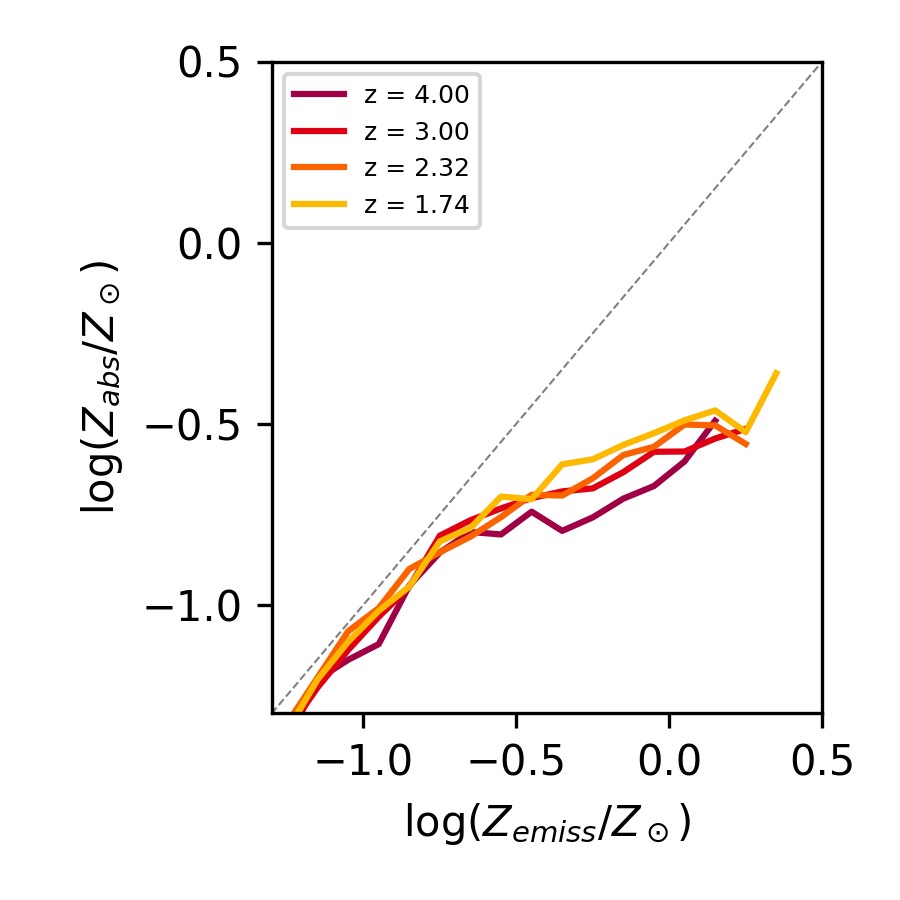}
	\includegraphics[width=0.49\textwidth]{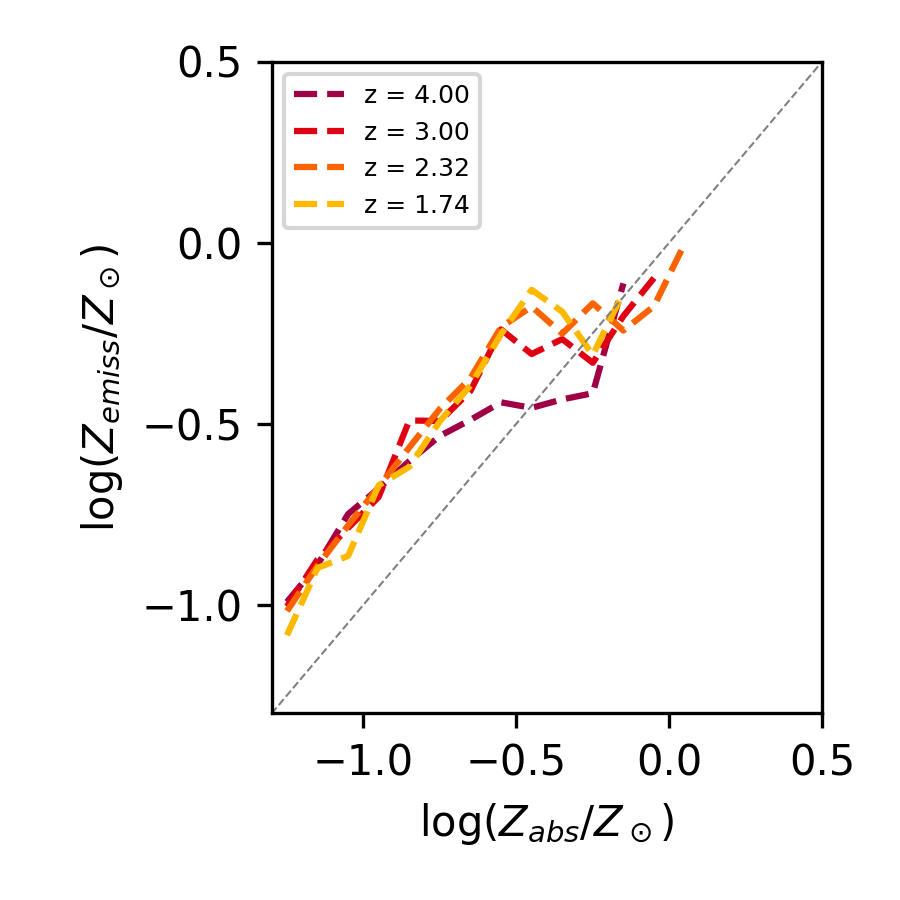}
    \caption{\emph{Left:} Redshift evolution of the median value of $Z_{\rm abs}$ for Illustris galaxies with a fixed value of $Z_{\rm emiss}$. \emph{Right:} Redshift evolution of the median value of $Z_{\rm emiss}$ for Illustris galaxies with a fixed value of $Z_{\rm abs}$. All of these plots assume a maximum metallicity for GRB progenitors of $Z_{\rm max}=0.4Z_\odot$.}
    \label{fig:OG_z-dep}
\end{figure*}

In Figure \ref{fig:OG_z-dep}, we show how the \Zabs-\Zem\ relation depends on redshift for the Illustris simulation, with a fixed value of $Z_{\rm max}=0.4Z_\odot$. While the median value of \Zabs\ at a fixed value of \Zem\ shows no significant redshift evolution, the median value of \Zem\ at fixed \Zabs\ shows a slight increase at lower redshifts, consistent with the results of the other two simulations. This trend can be attributed to the chemical enrichment of the universe between $z=4.0$ and $z=1.74$.

\section{Comparison between simulations} \label{sec:comparisons}

Several similarities can be seen between the \Zabs-\Zem\ relations predicted by all of the three cosmological simulations tested in this work. At fixed values of \Zem, the median value of \Zabs\ follows a particular shape; evolving near linearly until \Zem$\approx$\Zmax, and then turning off to a much shallower slope. For all simulations tested, this relation shows little evolution with redshift between $z=1.74$ and $z=4.0$. 

When \Zabs\ is fixed, the median value of \Zem\ does not depend on \Zmax, the maximum threshold for GRB production. Instead, the shape of this relationship depends on the populations of star forming galaxies present at each redshift, and how chemically homogeneous they are. For this reason, different simulations predict significantly different shapes for this relation. In EAGLE and TNG, this relationship depends substantially on redshift, with higher values of \Zem\ at lower redshifts; but for Illustris it is fairly constant over the redshift range we have examined.

Rather than consider the apparent tensions between different cosmological simulations as a weakness of our model, we see this as an opportunity. If different simulations predict different relationships between \Zem\ and \Zabs, then a set of observations of GRB host galaxy metallicities using both absorption and emission methods would hold the power to discriminate between different simulations of high-redshift galaxies using a set of data that none of these simulations have been tuned to reproduce. The \Zabs-\Zem\ relation is set by the overall degree of chemical inhomogeneity within galaxies. It depends not only on the presence of large-scale metallicity trends within these galaxies, but also on the small-scale structure of the gas inside galaxies, which varies on scales of $\sim 1$ kpc \citep[e.g.][]{GeoGals1, GeoGals2, Li+22}, and is difficult to observe outside of gravitationally-lensed targets at redshifts $\gtrsim 2$ \citep[e.g.][]{Treu+22, Wang+22}. Understanding the relationship between \Zem\ and \Zabs\ provides a novel set of data that allows us to observe chemical inhomogeneities in high-redshift galaxies from a new angle, giving new insights into understanding how feedback shapes galaxies at cosmic noon.

In Appendix \ref{ap:kdes}, we plot the full two-dimensional distributions of \Zabs\ and \Zem\ recovered for all galaxies, comparing each pair of simulations discussed in this work for every redshift and every value of \Zmax\ that we test. We find that regardless of the particular value of \Zmax, no two simulations make the same predictions for the \Zabs-\Zem\ relation at all redshifts. This means that, in principle, a survey that sought to measure \Zem\ of GRB host galaxies with known values of \Zabs\ at a variety of redshifts would be able to provide observational evidence for and against different models of galaxy formation and chemical evolution. We discuss this possibility further in Section \ref{sec:n_to_constrain_models}.

\subsection{What can GRB121024A tell us?}
\label{ssec:GRB121024}

At time of writing, only one GRB host galaxy exists for which both \Zem\ and \Zabs\ has been measured -- this is the host galaxy of GRB121024A \citep{Arabsalmani+18}. In this Section, we investigate what this data-point can tell us about the realism of the subresolution metal-enrichment schemes implemented in EAGLE, Illustris, and TNG.

This galaxy has a redshift of $z\sim 2.3$, with an absorption-metallicity of \logZabs$= -0.68 \pm 0.07$ \citep{Bolmer+19}, an emission-line metallicity of \logZem$ = -0.33^{+0.12}_{-0.24}$, and a Vega magnitude of $22.4 \pm 0.1$ in the K-filter of VLT/HAWK-I \citep{Friis+15}, corresponding to a K-corrected rest-frame U-band AB magnitude of $21.7 \pm 0.1$, assuming a flat SED \citep{ZZ_TNG}. To ensure a fair comparison to data, we follow \citet{ZZ_TNG} and include only simulated galaxies with a similar brightness to this galaxy, defined as those with rest-frame U-band magnitudes that are equal to this galaxy's, to within $0.75$. This magnitude cut is important, as it excludes a substantial number of galaxies with fainter magnitudes that would not be bright enough at this redshift for emission spectroscopy to be performed using current ground-based instruments. Therefore, without imposing a flux cut, the data-model comparison would be biased. %(Specifically X-shooter on the VlT; but do we care? 

For each simulation, the selection procedure described in Section \ref{sec:methods} was re-run after discarding all galaxies with rest-frame U-magnitudes outside of the selection range for each simulation's snapshot at $z=2.32$, in order to produce three magnitude-limited simulated samples of GRB120124A analogues. We note that this cut effectively limits the sample of host galaxies to only the brightest, most massive galaxies in each simulation, which do not generally follow the overall trends seen in the more general galaxy populations shown in the previous Sections.

The likelihood of each simulation hosting a bright GRB-host galaxy with similar values of \Zabs\ and \Zem\ to the host of GRB121024A was determined using the general method described in Section 4.1 of \citet{ZZ_TNG}. Briefly, for each galaxy in the magnitude-limited sample at $z=2.32$ of each simulation, the probability of the measured values of \Zabs\ and \Zem\ for the host galaxy of GRB121024A being drawn given (i) the uncertainties in the measured values of \Zabs\ and \Zem\ for the host of GRB121024A, and (ii) the simulated values of \Zabs\ and \Zem\ for each simulated galaxy, was computed. Then, this value was multiplied by the probability of each galaxy in the sample hosting a GRB given that one GRB is observed from a bright galaxy at $z=2.32$. The distribution for the true value of \Zabs\ for the host of GRB121024A was taken to be a log-normal distribution, as this is the highest entropy distribution for $\log($\Zabs$)$ when only the mean and uncertainty of this value are known \citep{Hogg+10}. For \Zem, the distribution was taken to be the posterior distribution as computed using NebulaBayes in \citet{ZZ_TNG}.

To thoroughly investigate the dependence of this result on the value of \Zmax, a finer grid was used for this investigation, exploring values of \Zmax\ ranging from $0.1Z_\odot$ to $1.0Z_\odot$ in increments of $0.1Z_\odot$, as well as the model with no cutoff metallicity. 
We found that the most likely model for formation of a GRB121024A host-analog galaxy was the Illustris simulation with a metallicity cutoff of \Zmax$=0.2Z_\odot-0.3Z\odot$.\footnote{The log-odds ratio between these two models is $-0.079$, which corresponds to a slight, nonsignificant ($\sim 20\%$) preference for a model with \Zmax$=0.3Z_\odot$. } The relative log likelihood of each other model compared to the most likely model is reported in Table \ref{tab:likelihoods}. At any fixed value of \Zmax, the most likely simulation to host a GRB121024A analog was Illustris. All of the EAGLE models are disfavoured, with log-odds ratios $\geq -1.3$, which corresponds to a significant $\geq 95\%$ preference for Illustris over EAGLE -- however, we note that this result is due to a dearth of bright galaxies produced in EAGLE at this redshift with values of \Zem\ similar to the host of GRB121024A and not the measured value of \Zabs. The most likely Illustris models are also favoured slightly over the most likely TNG model with \Zmax$=0.3Z\odot$ with a log-odds ratio of $\sim -0.3$, corresponding to a $\sim 2:1$ preference for the \Zabs-\Zem~ relation of Illustris over TNG. This result hints that the feedback procedures in Illustris may lead to more realistic chemical enrichment than in the TNG simulation, but this single data point is far from significant. A larger sample of observed data points would be required in order to make any definite claims about the relative realism of the feedback processes between these two models.

\begin{table}
\centering
\begin{tabular}{lrrr}
\hline
\Zmax &    \textbf{TNG} & \textbf{Illustris }& \textbf{ EAGLE} \\
\hline
\textbf{0.1}   & -0.703 &    -0.559 & -1.337 \\
\textbf{0.2  } & -0.450 &    -0.079 & -1.427 \\
\textbf{0.3 }  & -0.344 &    0      & -1.454 \\
\textbf{0.4 }  & -0.497 &    -0.180 & -1.382 \\
\textbf{0.5 }  & -0.574 &    -0.337 & -1.584 \\
\textbf{0.6 }  & -0.685 &    -0.404 & -1.495 \\
\textbf{0.7 }  & -0.806 &    -0.515 & -1.530 \\
\textbf{0.8 }  & -0.711 &    -0.518 & -1.546 \\
\textbf{0.9 }  & -0.845 &    -0.531 & -1.611 \\
\textbf{1.0}   & -0.842 &    -0.562 & -1.533 \\
\textbf{No Cutoff} & -1.291 & -0.937 & -2.021 \\
\hline
\end{tabular}
\caption{Relative log likelihoods (base 10) for each simulation/model hosting an analog of GRB121024A.}
\label{tab:likelihoods}
\end{table}

\begin{figure*}
	\includegraphics[width=0.97\textwidth]{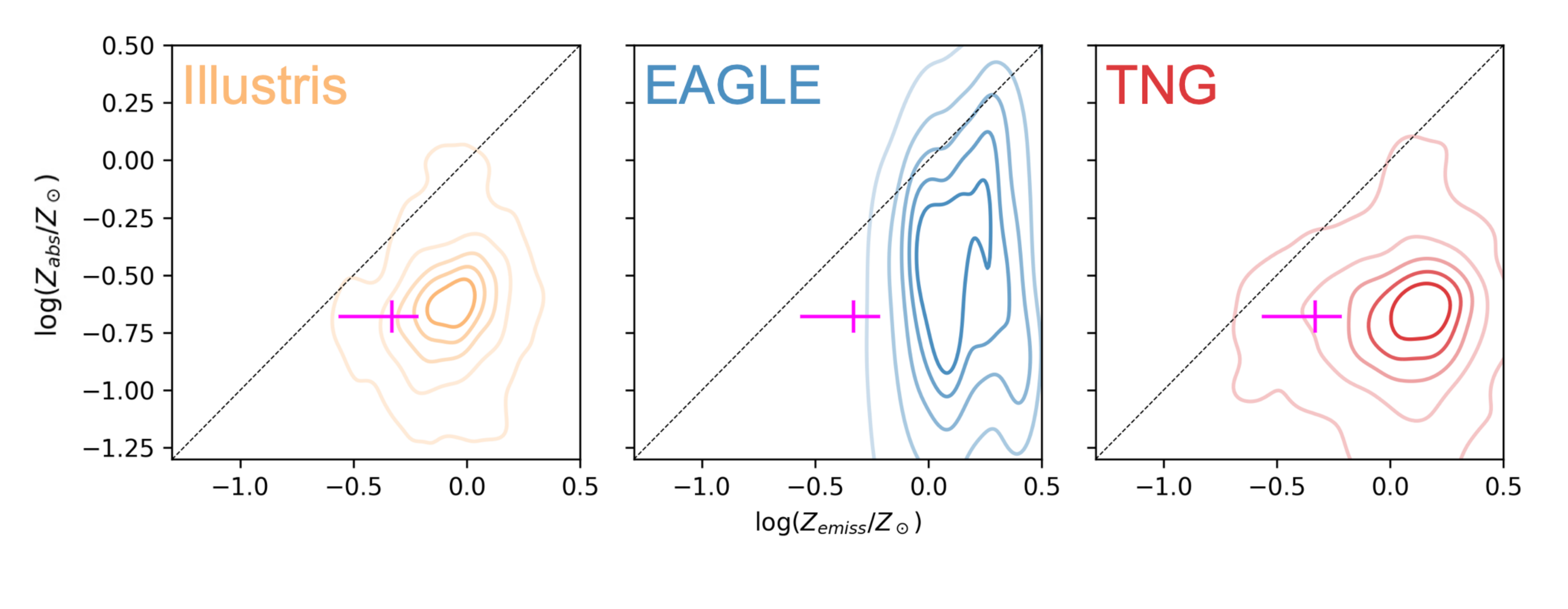}
    \caption{KDE plots showing the expected \Zabs-\Zem\ relation at $z=2.32$ for the magnitude-limited samples of simulated GRB hosts, assuming a value of \Zmax=$0.3Z_\odot$. In magenta, the location of the host galaxy of GRB121024A on the \Zabs-\Zem\ plane is plotted, with uncertainties. This data point shows the greatest agreement with the sample of bright simulated galaxies in Illustris, and disfavours the EAGLE simulation.}
    \label{fig:maglimited_KDEs}
\end{figure*}

To illustrate this point, in Figure \ref{fig:maglimited_KDEs}, we plot kernel density estimates for the magnitude-limited GRB host galaxy samples from each simulation at $z=2.32$, assuming \Zmax=$0.3Z_\odot$, which is one of the most likely models.\footnote{A metallicity cutoff of \Zmax$\approx 0.3Z_\odot$ is also supported by the investigations of \citet{Metha+20}.} We also plot in magenta the location of the \Zabs\ and \Zem\ measurements for the host of GRB121024A, too, so that this data point can be directly compared to the predictions of simulations. This plot shows that, while this data point agrees best with the predictions of the Illustris simulation and disfavours the EAGLE simulation, any of the three simulations tested would be capable of generating this data point, due to the large uncertainties on the measured values of \Zabs\ and \Zem.

\subsection{How many observations are needed to get a significant conclusion about the more likely feedback model?} \label{sec:n_to_constrain_models}

Using our simulated galaxy samples, we now consider the question: how many data points would need to be observed in order to provide a significant log-odds ratio that can be used to rule out certain simulation-driven models of the \Zabs-\Zem\ relation?

To assess this, the following procedure was undertaken. For each simulation, using each value of \Zmax, a GRB host galaxy was chosen based on its rate of GRB formation. Gaussian random noise was added to the logarithm of the values of \Zabs~ and \Zem~ for the simulated galaxy to reflect the difficulty in observing these quantities. For \Zabs, the observational uncertainty was modelled as being $0.11$ dex, which is the median value of the measured uncertainties of absorption-line based metallicity measurements of GRB host galaxies reported in \citet{ZZ_TNG}. For \Zem, the observational uncertainty was modelled as $0.2$ dex, typical of the uncertainties associated with strong emission-line based metallicity measurements at this redshift range \citep{Arabsalmani+18}.\footnote{This analysis does not account for any bias in the measured value of \Zem. Different emission-line ratio based metallicity diagnostics are known to produce metallicities that can vary in their absolute values by up to $0.7$ dex \citep[e.g.][]{Kewley+Ellison08}, so this bias may be significant. For the purposes of this analysis, we assume an observer that is successful at eliminating any bias from their metallicity measurements.} This data point (with its observational error) was then used to compute log likelihood ratios for the probability of generating this data point using the simulation that it came from over the probability of generating it from either of the two alternative models (using the same metallicity bias model for GRB progenitors every time). %For this analysis, the prior probability of each simulation was set to be equal.

This procedure was repeated $2.5 \times 10^{5}$ times, using all four downloaded snapshots at different redshifts for this comparison. The relative likelihood of a GRB coming from each redshift was taken using the observed cosmic GRB rate of SHOALS \citep{SHOALS1}. No magnitude-dependent selections were used to restrict the samples in this analysis, as this would depend on the capabilities of the survey instrument. 

\begin{figure}
	\includegraphics[width=0.45\textwidth]{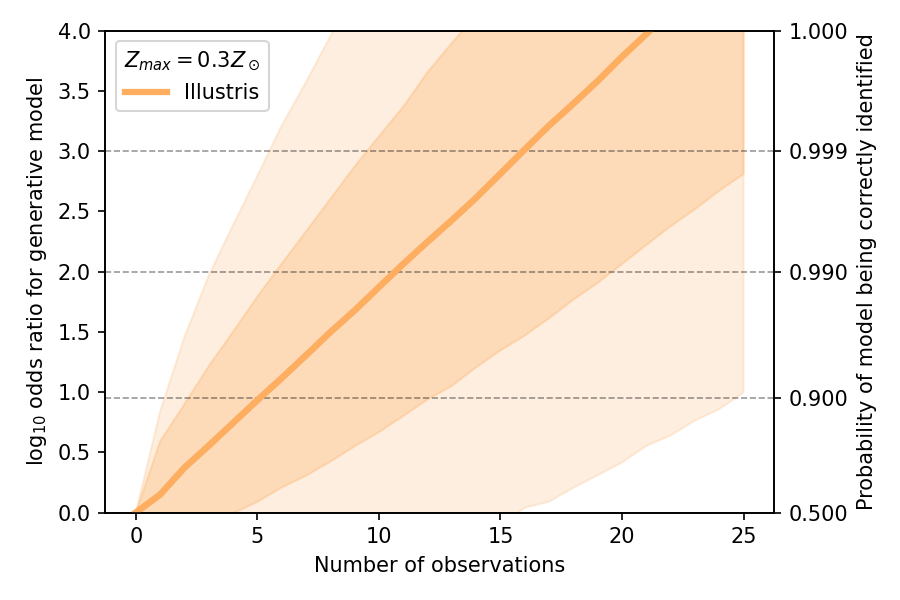}
	\includegraphics[width=0.45\textwidth]{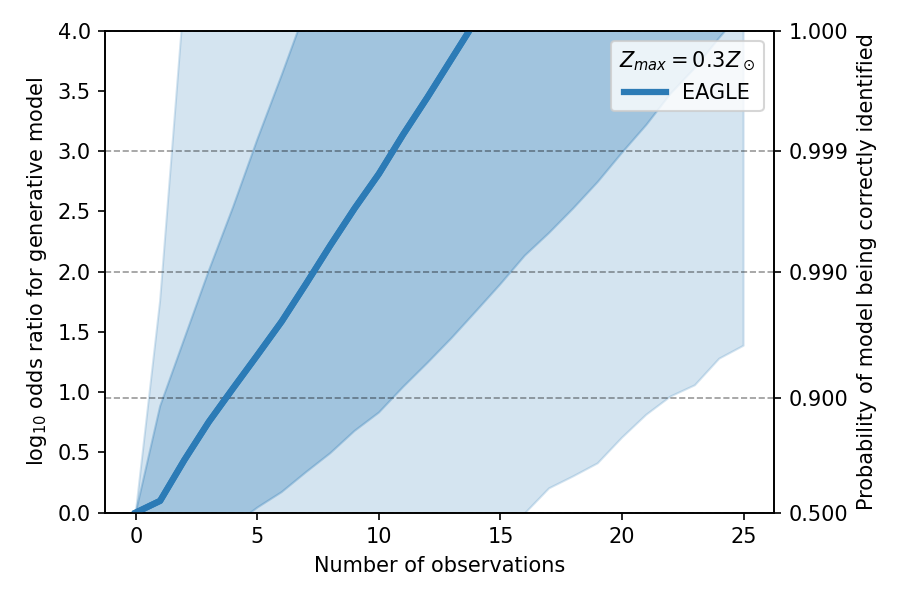}
	\includegraphics[width=0.45\textwidth]{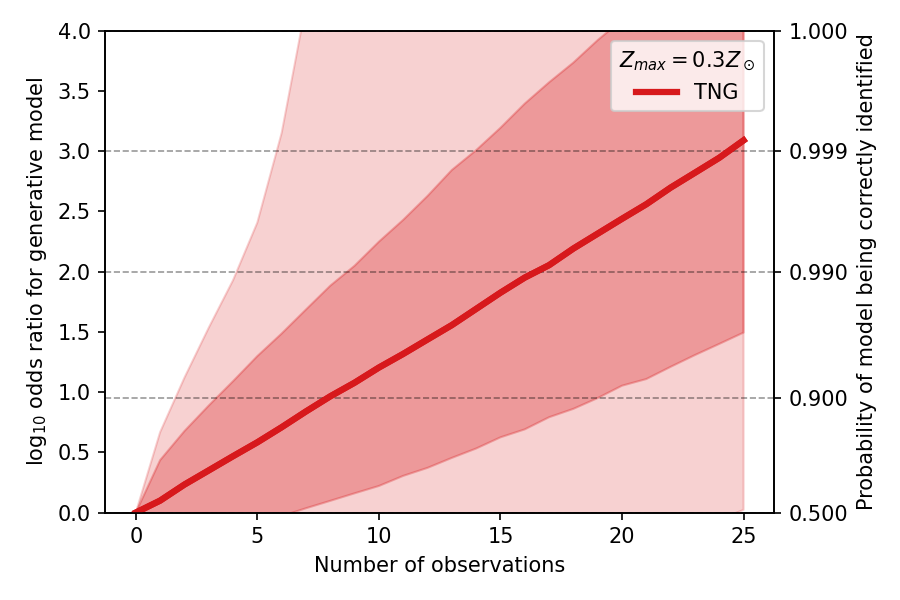}
    \caption{Odds ratio for the generative model as a function of the number of GRB host galaxies with observed values for \Zabs\ and \Zem. Bold lines represent the median log odds ratio, with the dark- and light-shaded regions representing the $68\%$ and $95\%$ symmetric confidence intervals around this median. We find that a small number $(N \sim 8-17)$ of observations is sufficient for distinguishing the correct model from the alternatives most of the time; however, the variance in the recovered log odds ratios were found to be very high, due to the width of the intrinsic scatter in the simulated \Zabs-\Zem\ relations and large observational uncertainties.}
    \label{fig:n_for_significance}
\end{figure}

These trials were organised into $10,000$ ``mock observational" runs, each observing values of \Zabs\ and \Zem\ for 25 GRB hosts at a redshift range of $z=1.74-4.0$. The increase in the evidence for the generative model (log-odds ratio) as the number of observations ($N$) increased was computed for each simulated observing run.
We show the median value of this relationship, and the 1-$\sigma$ and 2-$\sigma$ scatter, in Figure \ref{fig:n_for_significance} using our most likely model with \Zmax$=0.3Z_\odot$. We find that the EAGLE model of chemical enrichment and mixing generates a predicted \Zabs-\Zem\ relation that can be easily distinguished from the other two models. In over half of the trials conducted, only $N=8$ observations are needed in order to rule out the other models with high significance (a log-odds ratio greater than two). The Illustris simulation is the second easiest to separate, requiring about $N=11$ observations to reach this level of significance over $50\%$ of the time, with TNG requiring $N=17$ observations. This result can be explained by comparing the similarity of the subgrid physical prescriptions governing the structure of the ISM within these simulations. As the subgrid prescriptions and numerical methods used in the EAGLE simulation are very different to those used in the Illustris/TNG simulation, only a small number of observational data points are needed to discern the \Zabs-\Zem\ relation predicted by EAGLE from the other two alternative models. On the other hand, the TNG and Illustris simulations are far more similar. Furthermore, TNG has some feedback modes inspired by EAGLE (discussed in Section \ref{ssec:TNG}), which may lead to the internal chemical structures of star-forming galaxies in TNG being more similar to EAGLE than those in Illustris, increasing the number of observational data points needed to confidently conclude that the TNG model is the preferred generative model.

In all cases, the scatter in the possible values of the log odds ratio for the generative model over the alternatives as a function of the number of observed data points is very large. This is due to the intrinsic scatter within the \Zabs-\Zem\ relation predicted by all models (see Figures \ref{fig:E_v_I_KDEs}, \ref{fig:E_v_T_KDEs}, \ref{fig:I_v_T_KDEs}). For any generative models, it is possible to select either (i) a GRB host with values of \Zabs\ and \Zem\ that cannot be generated from another simulation, which vastly raises the likelihood, or (ii) a GRB host that is far more likely to come from one of the alternative models than the generative model, which lowers the odds ratio for identifying the correct simulation. Together, these two effects produce a very large variance in the log-odds ratio for each model as a function of $N$, the number of observations. Samples of GRB host galaxies drawn from the tails of the \Zabs-\Zem\ distribution are necessary to distinguish between the models investigated in this work. This may be achieved in as few as $N=5$ observations, or it may take more than $N=25$.
By examining the two-dimensional KDEs for the three simulations compared in this work (Appendix \ref{ap:kdes}), we see that targeting low-redshift galaxies with 
low values of \Zem\ 
%a range of \Zabs\ values
is the best way to distinguish between the different predicted \Zabs-\Zem\ relations. However, in practise, the number of GRBs with measured values of \Zabs\ for which spectroscopic follow-up is feasible is limited; and the exact number of GRB hosts that are needed to constrain this relation is simply up to chance.

This analysis was repeated for models other than the most likely model with \Zmax$=0.3Z_\odot$. We found that for lower values of \Zmax\ (stricter GRB metallicity bias functions), a smaller number of data points were needed to correctly identify the generative model in all cases. When GRBs were assumed to be unbiased tracers of star formation, a median value of $15-20$ observations were required to achieve a log-odds ratio of $1$ for the correct generative model. This indicates that using metallicity-biased tracers of star formation, such as GRBs, can play an important role in determining the internal metallicity structure of distant galaxies.

\section{Discussion} \label{sec:disco}

As discussed in Section \ref{ssec:GRB121024}, currently only one GRB host galaxy has had its metallicity measured with both strong emission line diagnostics and by examining the absorption spectrum of the transient. A significant difference is seen between \Zabs\ and \Zem\ ($0.35^{+ 0.14}_{- 0.25}$ dex; \citealt{ZZ_TNG}). Similar differences have also been reported by various teams examining the relationship between \Zabs\ and \Zem\ for quasar-absorbing galaxies: \citet{Christensen+14} report a mean offset of $0.44 \pm 0.10$ dex. However, the physical origin of this offset is very different for quasar- and GRB- absorbing systems.

For quasar-absorbing systems, \citet{Christensen+14} find that the difference between the measured values of \Zabs\ and \Zem\ correlates heavily with the distance between the quasar line-of-sight and the galaxy centre. Such a relationship is expected from any theory of galaxy formation that produces negative metallicity gradients \citep[e.g.][]{Boissier+Prantzos99}. As \Zem\ is biased to the brightest regions of star formation and galaxies are generally brightest in their central regions, \Zem\ can be taken as a tracer of the metallicity near the centre of a galaxy, while \Zabs\ traces only the metallicity of the gas along the line-of-sight. An offset between \Zem\ and \Zabs\ is expected, then, when a quasar line-of-sight passes through the outskirts of a galaxy, due to the different environments that these two processes are probing.

On the other hand, for GRB-absorbing systems, the line-of-sight necessarily originates from a region of recently star-forming gas, as GRBs are associated with core collapse supernovae which have short delay times \citep{Fruchter+06, Briel+22}. For this reason, the spatial regions traced by \Zabs\ and \Zem\ for GRB host galaxies must overlap, to some extent. Furthermore, at high redshifts, galaxies with a wide variety of metallicity gradients are seen, both in observations \citep[e.g.][]{Cresci+10, Leethochawalit+16, Wuyts+16, Wang+22} and in simulations \citep{Ma+17, Hemler+21, Tissera+22}. Therefore, we must carefully consider how \Zabs\ is expected to differ from \Zem\ in galaxies with negative (standard) metallicity gradients, positive (inverted) gradients, and galaxies with no metallicity gradient at all. We discuss each of these cases below.

If a galaxy has a negative metallicity gradient, then \Zabs\ is expected to be lower than \Zem. Under our model, only low-metallicity regions of a galaxy are able to form GRBs. This means it is more likely for a GRB to form away from the high-metallicity central regions of these galaxies. While it is still possible for the line-of-sight from this GRB to pass through the higher metallicity central regions, it is far more likely for it to go either out of the disc, or pass through the disc away from the central regions, avoiding the high metallicity central regions. On the other hand, \Zem\ will be biased to the metallicity of the central region where the star formation rate is highest, making \Zem\ larger than \Zabs.

If a galaxy has an inverted/positive metallicity gradient, then \Zabs\ is still expected to be higher than \Zem, on average, but with a larger scatter compared to the previous case. In fact, in this scenario, GRB formation will be biased to the low-metallicity central regions of a galaxy. This will especially be true if the inverted metallicity gradient comes from a cold gas inflow into a galaxy's central regions, which would trigger enhanced star formation in the low-metallicity centre \citep[as suggested by e.g.][]{Dekel+09a, Dekel+09b, Stott+14, Ellison+18}. When the GRB line-of-sight is directed out of the disc, \Zabs\ will reflect predominately the low metallicity of the galaxy's central regions. This will be lower than \Zem, which will be integrated over a larger area including higher metallicity regions. However, when the line-of-sight of the GRB is aligned such that it passes through the disc, \Zabs\ will return the integrated metallicity along higher metallicity regions. Depending on the orientation of the jet, this may result in a value of \Zabs\ that is comparable to or larger than \Zem. % \textbf{OoM calculation here about a disc flare angle?}

For a galaxy with no metallicity gradient, the relationship between \Zabs\ and \Zem\ depends on how well metals are mixed throughout the galaxy. For a perfectly homogeneous galaxy, \Zem\ and \Zabs\ would be equal. However, recent studies of the two-dimensional metallicity structure of local galaxies have shown that they exhibit significant variations in metallicity (of the order of $\sim 0.1$ dex) on scales of $\sim 1$ kpc \citep{GeoGals1, GeoGals2, Li+22}. Such small-scale variations in metallicity may allow GRBs to be formed in low-metallicity regions of high-metallicity galaxies, along lines of sight that, when integrated, allow values of \Zabs\ that are still lower than \Zem.

All of these results suggest that the \Zabs-\Zem\ relation does not trace whether or not galaxies have negative, positive, or no gradients. Rather, the shape of this relationship depends on the overall degree of chemical inhomogeneity within galaxies at high redshift. As the numerical methods and subresolution physical prescriptions of all these simulations are very different (see Section \ref{sec:simulations}), it would be very challenging to ascribe differences in the \Zabs-\Zem\ relations predicted between these simulations (which depends on the small-scale chemical structure of the ISM of the simulated galaxies) to any specific features, such as the feedback models used, or the presence or absence of magnetism. All factors described in Section \ref{sec:simulations} likely play a role, to some extent. A rigorous discussion into what sets the \Zabs-\Zem\ relation is beyond the purpose of this paper, as our focus is more on observational consequences.

Qualitatively, the form of the \Zabs-\Zem\ relation is consistent among a selection of three galaxy formation models (Figures \ref{fig:TNG_z=2.33_median}, \ref{fig:EAG_z=2.33_median},  \ref{fig:OG_z=2.33_median}). 
For all simulations the most likely value of \Zmax\ was fit using the data on the host galaxy of GRB121024A, and was found to be $0.3Z_\odot$ for Illustris and TNG, and could only be constrained to $\leq 0.4 Z_\odot$ for EAGLE. 
The agreement between these different simulations in determining this value tells us that we may be confident in using the \Zabs-\Zem\ relation to determine \Zmax\ robustly. We note that for the EAGLE simulation only galaxies with \Zem$ \gtrsim 0.5 Z_\odot$ show differences between the median value of \Zabs\ at fixed \Zem\ for different values of \Zmax. 

In \citet{ZZ_TNG} it was shown that using the TNG simulation, the median value of \Zem\ at a specific \Zabs\ depends not on the value of \Zmax. This result was also found in this work when EAGLE and Illustris were used. However, as was conjectured in \citet{ZZ_TNG}, the dependence of the value of \Zem\ at fixed \Zabs\ was found to be very sensitive to the model of galaxy evolution employed.

This means that the \Zabs-\Zem\ relation can tell us two things: (i) the metallicity bias function for GRB progenitor stars, and (ii) some information about how metals are mixed and redistributed in high-redshift galaxies. This information can be used to compare cosmological simulations along a new axis that has not been calibrated \textit{a priori}, to provide a novel test that is sensitive to the sub-resolution models of chemical enrichment and galaxy evolution.

\section{Summary and conclusions}
\label{sec:conc}

In this study, we investigated the impact that chemical inhomogeneities of star-forming galaxies have on the relationship between \Zabs, the metallicity measured using absorption-line spectroscopy, and \Zem, the metallicity measured using gas emission lines, for GRB host galaxies. We analysed three distinct hydrodynamical cosmological simulations: EAGLE, Illustris, and IllustrisTNG, all of which employ different numerical methods and subgrid physical prescriptions to generate realistic populations of galaxies. Using a consistent methodology across all three simulations, GRB host galaxies were identified from snapshots at different redshifts through a Monte Carlo approach that includes a preference for low metallicity hosts (treated as a free parameter). The relationship between \Zabs\ and \Zem\ was computed using a range of GRB metallicity bias functions at redshifts $1.74\leq z \leq 4.0$. Our main conclusions are as follows:

\begin{itemize}
    \item Different simulations predict different forms of the \Zabs-\Zem\ relation. This implies that this relationship, which depends on the degree of chemical homogeneity within galaxies, is sensitive to choices of sub-resolution physics and details of chemical enrichment and mixing used in simulations.
    \item The form of the \Zabs-\Zem\ relation predicted from the EAGLE simulation is notably different to that predicted from the Illustris and TNG simulations. At all redshifts and for all values of \Zmax\ tested, the values of \Zabs\ and \Zem\ are more likely to agree for this simulation. This implies that most high-redshift galaxies in EAGLE are predicted to be more homogeneous than those produced in Illustris or TNG.
    \item On the other hand, the EAGLE simulation also exhibited a small population ($\sim 5-10\%$) of galaxies with extremely low values of \Zabs\ ($10^{-5} - 10^{-2} Z_\odot$). These galaxies have regions of extremely metal poor gas, indicating that the feedback and metal mixing schemes implemented in EAGLE may be inefficient at redistributing metals into a galaxy's outer regions. Such a low \Zabs\ GRB host population was not seen in either Illustris or TNG.
    \item A statistical analysis of the \Zem\ and \Zabs\ values determined for the host galaxy of GRB121024A was constructed. This single data point favours the TNG and Illustris models with a metallicity cutoff at \Zmax$\approx 0.3Z_\odot$, and specifically prefers Illustris over TNG. The EAGLE simulation was significantly disfavoured, as bright galaxies with low values of \Zem$<0.5Z_\odot$ are rare in this simulation.
    \item Using Monte Carlo methods, a mock survey was constructed measuring the \Zabs-\Zem\ relation across a  redshift range of $1.74-4.0$ for the three simulations. We found that, assuming a GRB bias function with \Zmax$=0.3Z_\odot$, an observing campaign targeting $8-17$ GRB hosts would likely be sufficient to gather a significant amount of evidence for or against the modelled metallicity distributions of these high-redshift simulated galaxies. 
\end{itemize}

An upcoming observational campaign using the JWST will collect \Zem\ measurements for ten GRB host galaxies at $2.0<z<4.7$ for which \Zabs\ has already been determined \citep{Schady_JWST}. Data from this population of GRB host galaxies will greatly increase the constraining power of observations on the \Zabs-\Zem\ relation. In fact, we expect the dataset will allow us to achieve a better understanding of both the metallicity bias for GRB progenitors, and the small-scale metallicity structures of high-redshift galaxies. % If we are fortunate, some of these GRB hosts may occupy regions of the \Zabs-\Zem\ plane that are inconsistent with the predictions of any or all of these simulations, allowing us to gain deeper evidence for or against the different models of stellar feedback used by these simulations.
However, based on our analysis, whether or not this sample of $N=10$ additional GRB host galaxies will be large enough to discriminate between any of the \Zabs-\Zem\ relations predicted from different cosmological simulations at a reasonably high confidence level still remains to be seen. In fact, our results shown in Figure~\ref{fig:n_for_significance} indicate that the sample size may have to be increased to $N\gtrsim17$. Given that the limiting factor for further progress is the availability of GRBs with robust absorption metallicity measurements at high redshift, future facilities for all-sky gamma-ray monitoring and for rapid follow-up at infrared wavelengths such as the HERMES/SpIRIT constellation (\citealt{Fiore+20}, Thomas et al. 2022, submitted) and the SkyHopper infrared space telescope \citep{Thomas+22} have the potential to play a contributing role along with flagships such as JWST to advance our understanding of how chemical elements are distributed in the interstellar and intergalactic medium.

\section*{acknowledgements}
The authors thank the anonymous referee for providing insightful comments and suggestions that improved the quality of the manuscript. BM acknowledges support from an Australian Government Research Training Program (RTP) Scholarship and a Laby PhD Travelling Scholarship. This research is supported in part by the Australian Research Council Centre of Excellence for All Sky Astrophysics in 3 Dimensions (ASTRO 3D), through project number CE170100013.

\section*{data availability}

Data products from the EAGLE simulation suite are available for public download via the Virgo consortium's website: \url{virgodb.dur.ac.uk}. Data products from the Illustris simulation are available from \url{www.illustris-project.org/data/}. Data products from the IllustrisTNG simulation are available from \url{www.tng-project.org/data/}. 
Any further data products presented in this work, as well as the Python code used to construct them, are available from the corresponding author (BM) upon reasonable request.

\bibliographystyle{mnras}
\bibliography{biblio} 

\newpage

\appendix

\section{Comparison of \Zabs - \Zem relations between simulations}
\label{ap:kdes}

\begin{figure*}
	\includegraphics[width=0.95\textwidth]{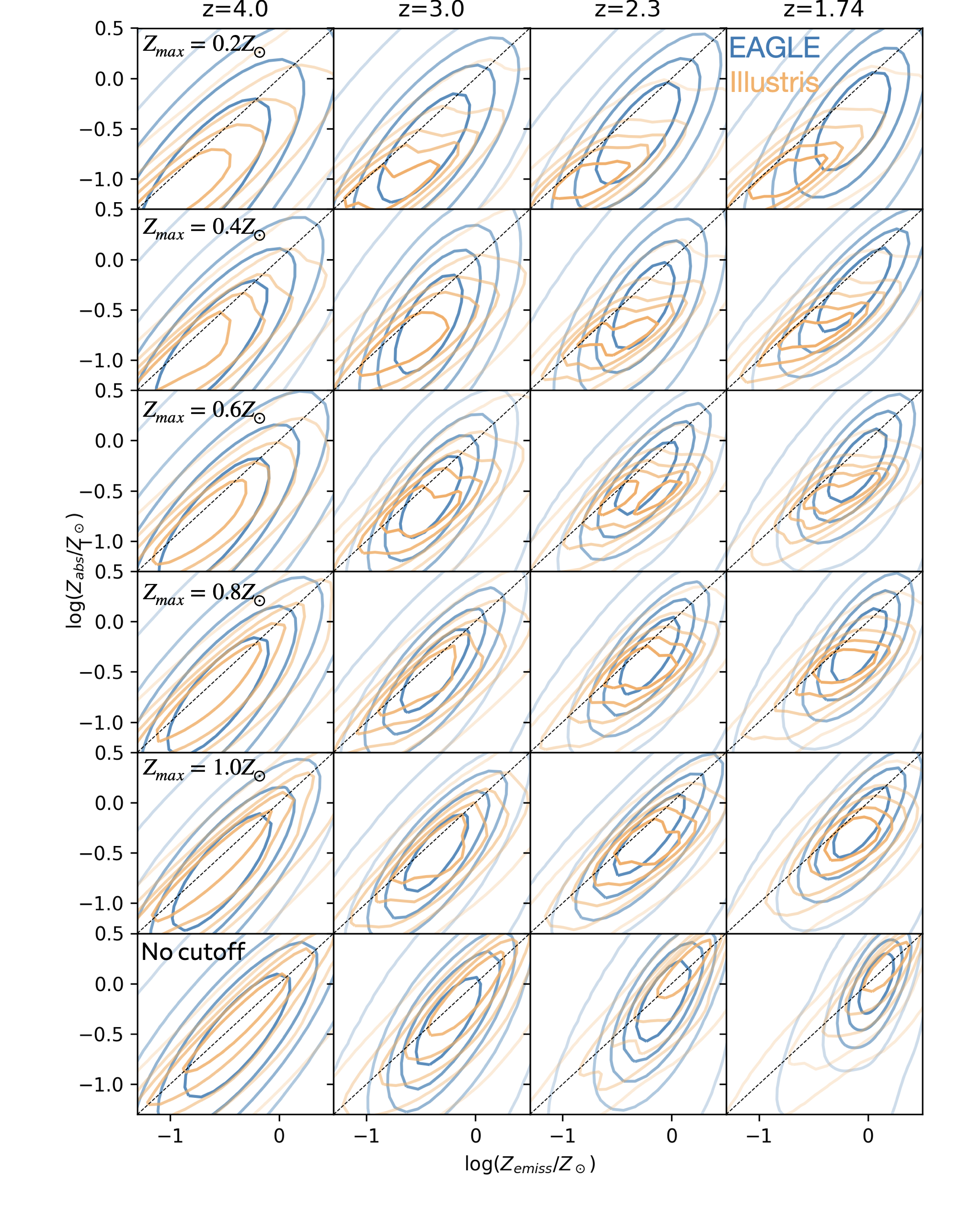}
    \caption{Comparing the predicted \Zabs-\Zem\ relations between the Illustris and EAGLE simulations, at all redshifts, for all GRB bias models tested.}
    \label{fig:E_v_I_KDEs}
\end{figure*}

\begin{figure*}
	\includegraphics[width=0.95\textwidth]{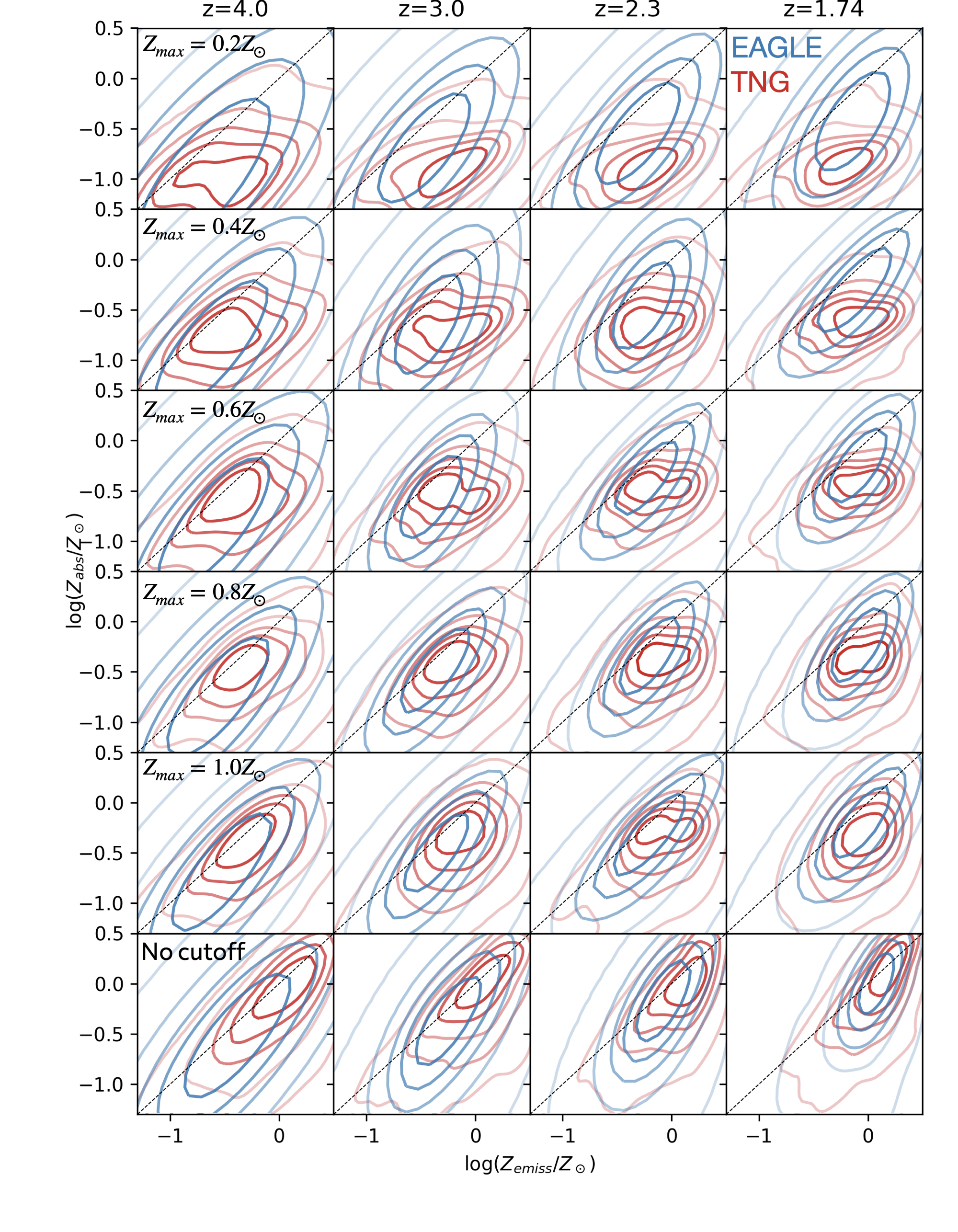}
    \caption{Comparing the predicted \Zabs-\Zem\ relations between the EAGLE and TNG simulations, at all redshifts, for all GRB bias models tested.}
    \label{fig:E_v_T_KDEs}
\end{figure*}

\begin{figure*}
	\includegraphics[width=0.95\textwidth]{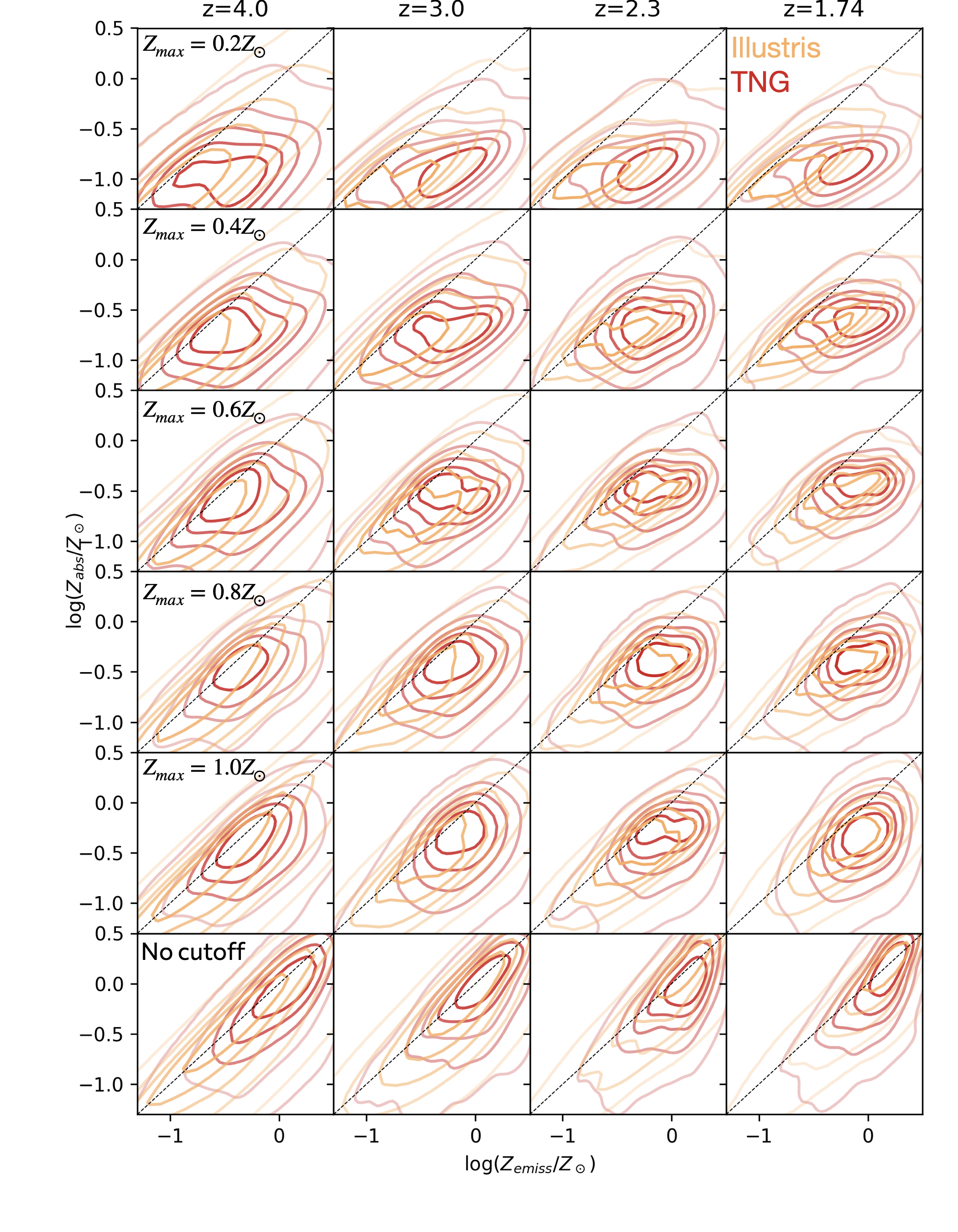}
    \caption{Comparing the predicted \Zabs-\Zem\ relations between the Illustris and TNG simulations, at all redshifts, for all GRB bias models tested.}
    \label{fig:I_v_T_KDEs}
\end{figure*}

In Figures \ref{fig:E_v_I_KDEs}, \ref{fig:E_v_T_KDEs}, and \ref{fig:I_v_T_KDEs}, we compare the predicted \Zabs-\Zem\ relations between each pair of simulations, showing the full two-dimensional kernel density estimates for each simulation, varying the redshift and the metallicity cutoff for GRB formation. 

From these plots, it is clear that the predicted \Zabs-\Zem\ relation depends greatly on the subgrid prescriptions of the model used. The differences between the predicted simulations becomes more severe at lower values of \Zmax, as details of the internal metallicity structure of galaxies have a greater role in setting the morphology of the \Zabs-\Zem\ relation when a more stringent cutoff is used. Differences seen are also more intense for low redshift. Here, metallicities of galaxies are generally higher, so imposing a low metallicity cutoff has a greater effect on the \Zabs-\Zem\ relation. This is good news for prospects of future observations, as \Zem\ is easier to measure at lower redshift.

These plots also demonstrate how both \Zabs\ and \Zem\ evolve with redshift. In all simulations and for all values of \Zmax\ displayed, at lower redshifts, values of both \Zabs\ and \Zem\ are higher. This reflects the chemical enrichment of the universe, which all simulations capture. Note that this result does not contradict our result of Section \ref{sec:results}, in which we showed that the median value of \Zabs\ at a fixed value of \Zem\ shows very little redshift evolution in any simulation (see left hand panel of Figures \ref{fig:TNG_z-dep}, \ref{fig:EAGLE_z-dep},  \ref{fig:OG_z-dep}): at higher redshifts, the number of galaxies with high values of \Zem\ is lower; but for any fixed value of \Zem, the median value of \Zabs\ shows very little dependence on redshift.

Even when the predicted distributions peak at the same location on the \Zabs-\Zem\ plane, different models show large differences in the tails of the simulations, implying that with a large enough collection of observations, a few rare events may be able to distinguish between such models. This implies that a survey aimed at collecting values of \Zem\ for GRB host galaxies for which \Zabs\ estimates have already been established at a range of redshifts will be able to constrain models of chemical evolution within galaxies (see Section \ref{sec:n_to_constrain_models} for further details).

\label{lastpage}
\end{document}